\documentstyle[]{mn} 
\input epsf
\input psfig.sty
\def\gsim{\lower.4ex\hbox{$\;\buildrel >\over{\scriptstyle\sim}\;$}}
\def\lsim{\lower.4ex\hbox{$\;\buildrel <\over{\scriptstyle\sim}\;$}}

  \def\bib{\bibitem{}}
\newcommand{\xia}{\overline{\xi}}
\newcommand{\rhoa}{\overline{\rho}}
\newcommand{\gam}{\gamma}
\newcommand{\Na}{\overline{N}}
\newcommand{\na}{\overline{n}}
\newcommand{\xit}{\tilde{\xi}}
\newcommand{\sigmat}{\tilde{\sigma}}
\newcommand{\Ft}{\tilde{F}}

\newcommand{\beq}{\begin{equation}}
\newcommand{\eeq}{\end{equation}}
\newcommand{\lag}{\langle}
\newcommand{\rag}{\rangle}
\begin{document}
%
% US additions
%
\leftmargin= 2.5 cm
\renewcommand{\textfraction}{.01}
\renewcommand{\topfraction}{0.99}
\renewcommand{\bottomfraction}{0.99}
\setlength{\textfloatsep}{2.5ex}

\title{Scaling laws in gravitational clustering for counts-in-cells and mass functions}

\author[P. Valageas, C. Lacey and R. Schaeffer] {P. Valageas$^{1,2,3}$, C. Lacey$^{2}$ and R. Schaeffer$^{1}$\\
$^1$Service de Physique Th\'eorique, CEA Saclay, 91191 Gif-sur-Yvette, France\\
$^2$Theoretical Astrophysics Center, Juliane Maries Vej 30, 2100 Copenhagen O, Denmark\\
$^3$Center for Particle Astrophysics, Department of Astronomy and Physics, University of California, Berkeley, CA 94720-7304, USA}

\maketitle 

\begin{abstract}

We present in this article an analysis of some of the properties of
the density field realized in numerical simulations for power-law
initial power-spectra in the case of a critical density universe. We
study
the non-linear regime, which is the most difficult to handle
analytically, and we compare our numerical results with the predictions
of a specific hierarchical clustering scaling model that have been
made recently, focusing specifically on its much wider range of
applicability, which is one of its main advantages over the standard
Press-Schechter approximation. We first check that the two-point
correlation functions, measured both from counts in cells and
neighbour counts, agree with the known analytically exact scaling
requirement (i.e. only depend on $\sigma^2$) and we also find the
stable-clustering hypothesis to hold. Next we show that the statistics
of the counts in cells obey the scaling law predicted by the above
scaling model.

Then we turn to mass functions of overdense and underdense regions,
that we obtain numerically from ``spherical overdensity'' and
``friends-of-friends'' algorithms. We first consider the mass function
of ``just-collapsed'' objects defined by a density threshold
$\Delta=177$ and we note as was found by previous studies that the
usual Press-Schechter prescription agrees reasonably well with the
simulations (although there are some discrepancies). On the other hand,
the numerical
results are also consistent with the predictions of the scaling
model. Then, we consider more general mass functions (needed to
describe for instance galaxies or Lyman-$\alpha$ absorbers) defined by
different density thresholds, which can even be negative.  The scaling
model is especially suited to account for such cases, which are out of
reach of the Press-Schechter approach, and it still shows reasonably
good agreement with the numerical results. Finally, we show that mass
functions defined by a condition on the radius of the objects also
satisfy the theoretical scaling predictions.

Thus, we find that the scaling model provides a reasonable description
of the density field in the highly non-linear regime, for the
cosmologies we have considered, both for the counts in cells
statistics and the mass functions. The advantages of this approach are
that it clarifies the links between several statistical tools and it
allows one to study many different classes of objects, for any density
threshold, provided one is in the fully non-linear regime.

\end{abstract}


\section{Introduction}

In the standard cosmological scenario large-scale structures in the
universe arise from the amplification through gravitational
instability of small primordial density fluctuations. These initial
perturbations are likely to be gaussian (as in most inflationary
models) and are characterized by their power-spectrum.  Within the
hierarchical clustering scenario, the amplitude of these fluctuations
increases towards the smaller scales (as in the CDM model: Peebles
1982; Davis et al.1985). Small scales collapse first to form bound
objects which merge later to build increasingly massive halos as
larger scales become non-linear. These mass condensations correspond
to the various astrophysical objects one can observe in the universe,
from Lyman-$\alpha$ clouds to galaxies and clusters. As a consequence,
to understand the formation of these objects it is important to obtain
a precise description of the evolution of the density field under the
action of gravity. Note that to model specific astrophysical objects
one usually needs to add to this description of dark matter halos
non-gravitational effects (such as, for instance, radiative processes
to get the evolution of baryons). This is beyond the scope of the
present paper that considers specifically the multiplicity of dark
halos.

However, even for the problem of the evolution of the matter
distribution under the sole action of gravity theoretical results are
very scarce. Linear theory allows one to describe the density field on
large scales, while a few approximations (e.g. see Bernardeau 1996) try
to handle the early non-linear evolution, but the highly non-linear
regime has proved very difficult to model. As a consequence numerical
simulations have so far been the main tool to describe this latter
stage. In this article, we compare the results of simulations for
power-law initial spectra in the case of a critical density universe
with the scaling model presented in previous publications (Valageas \&
Schaeffer 1997, hereafter VS; Bernardeau and Schaeffer 1991; Balian \&
Schaeffer 1989a ). This latter description of the density field is based
on the assumption that the many-body correlation functions satisfy
specific scaling laws (obtained from the stable-clustering ansatz) in
the highly non-linear regime. In this case one obtains a very powerful
model for the density field, which can be used to obtain the counts in
cells statistics as well as mass functions of overdense and underdense
regions at different density thresholds. Note that in contrast the
Press-Schechter (PS) mass function (Press \& Schechter 1974), which is
certainly the most popular tool to get some information on the
characteristics of the non-linear density field, only deals with
``just-collapsed'' objects.

The article is organized as follows. In Sect.\ref{Simulations} we
describe the numerical simulations we analyse. Then we present our
results for the two-point correlation function and the counts in cells
statistics in Sect.\ref{Counts-in-cells}. We compare the numerical
mass function of ``just-collapsed'' objects to the usual PS
prescription and to the scaling model in Sect.\ref{Mass functions}. We
also consider more general mass functions, beyond the reach of the PS
approach, that are defined by various density thresholds. Finally we
present our results for the limiting case of mass functions of objects
defined by a constant radius constraint. These studies originate from
astrophysical descriptions of galaxies (Valageas \& Schaeffer 1999; Valageas \& Silk 1999)
and Lyman-$\alpha$ clouds (Valageas et al.1999) where one is naturally
led to introduce such generalized mass functions.

\section{Simulations}
\label{Simulations}

We shall consider a critical density universe, $\Omega=1$, with an
initial
power-spectrum $P(k)$ which is a power-law: $P(k) \propto k^n$. We
study the cases $n=-2$, $n=-1$ and $n=0$. As usual we shall define
$\sigma^2(R,a)$ to be the amplitude of the density fluctuations in
cells of
physical radius $R$ at time $t$ (scale-factor $a$) given by linear
theory. Thus we have:
\beq
\sigma^2 \propto a^2 \; r^{-(n+3)} \propto a^{(n+5)} \; R^{-(n+3)}
\eeq
where $r=R/a$ is a comoving scale.

The N-body simulations were performed using the $AP^3M$ code of
Couchman (1991). The nominal box size was $L=256$ Mpc/h (though this
could be rescaled to any value, since the initial conditions are
scale-free). The simulations all used $N_p = 128^3 \approx 2\times
10^6$ particles, with a force-softening parameter (constant in
comoving coordinates) of $0.1$ times the mean interparticle separation
$L/N_p^{1/3}$, that is over a comoving radius of $0.2$ Mpc/h. 
The expansion factor $a$ was normalized so that
$\sigma(R,a)=1$ for $R=8$ Mpc/h at $a=1$, according to linear
theory. The initial positions and velocities of the particles were
given by displacing the particles from a cubical grid using the
Zeldovich approximation and the linear theory power spectrum. The
starting time was chosen small enough so that the density fluctuations
on the scale of the particle grid were still close to the linear
regime. Specifically, the linear power-spectrum amplitude at the
Nyquist frequency of the particle grid was chosen to be $A^2$ times
the white noise value, with $A=1$ for $n=0$ and $n=-1$, and $A=0.4$
for $n=-2$, corresponding to expansion factors $a_i=0.0611,0.15,0.194$
for $n=0,-1,-2$ respectively. For $n=0$, the simulation was evolved
with a constant timestep $\Delta p= 2.36\times 10^{-3}$ in the
variable $p = a^{2/3}$, while for $n=-1$ and $-2$ a constant timestep
in $a$ was used, with $\Delta a = 3.4\times 10^{-3}, 1.5\times
10^{-3}$ respectively. The simulations were evolved up to expansion
factors $a_f=8,4,2.67$ for $n=0,-1,-2$ respectively. In the numerical
analysis we shall use the output times in the range $1 \leq a \leq
a_f$ when some non-linear structures have already formed on scales
larger than the smoothening length.

For this cosmology and initial power spectra, the real clustering
evolution should be {\it self-similar}, when scaled to a radius $R_*(a)
\propto a^{(n+5)/(n+3)}$ such that $\sigma(R_*,a)=1$. This is an exact
analytical result that
holds independently of the validity of the Balian \& Schaeffer (1989a )
{\it scaling } predictions (see Sect.\ref{Counts-in-cells}) that we aim
at testing in this paper.
 The evolution in the
N-body simulation will depart from exact self-similarity because of
numerical effects, in particular, particle discreteness, force
softening, the finite box size, and the absence of initial
fluctuations smaller than the Nyquist wavelength or larger than the
box size.

\section{Counts-in-cells}
\label{Counts-in-cells}

A convenient way to describe the density field obtained in a numerical
simulation, or realized in the actual universe, is to consider the
counts in cells. Thus, we define the probability distribution
$P_R(N;t)$ to be the probability to have $N$ particles in a spherical
cell of radius $R$ at a given time $t$. In the following we shall
usually denote $P_R(N;t)$ as $P(N)$ to simplify the notation. This is a
well-defined quantity which provides a very good description of the
density field and is a convenient tool for a theoretical analysis. In
contrast, the multiplicity functions used to recover the counts of
virialized halos or astrophysical objects may be defined in various
ways and are somewhat more difficult to handle analytically. However,
in certain regimes their properties can be obtained from the
characteristics of the counts in cells. Hence we shall first consider
the statistics of $P(N)$ in the next sections.

\subsection{Non-linear scaling model}
\label{Scaling model}

Since we shall mainly compare our results with the scaling model
described in Balian \& Schaeffer (1989a ) we recall here their predictions
 while introducing our notation. This model is based on
the assumption that the many-body correlation functions $\xi_p({\bf
r}_1,...,{\bf r}_p;a)$ satisfy the scaling law :
\beq
\xi_p(\lambda {\bf r}_1,...,\lambda {\bf r}_p ;a) = a^{3(p-1)} \;
\lambda^{-\gam(p-1)} \; \hat{\xi}_p({\bf r}_1,...,{\bf r}_p)
 \label{scal1}
\eeq
where $a(t)$ is the scale-factor and $\gam$ is the slope of the
two-point correlation function (which we note $\xi$). For an initial
power-spectrum $P(k)$ which is a power-law, $P(k) \propto k^n$, we
have:
\beq
\gam = \frac{3(3+n)}{5+n}  \label{gam1}
\eeq
The relations (\ref{scal1}) and (\ref{gam1}) are derived from the
stable-clustering assumption (Davis \& Peebles 1977; Peebles 1980) and
are expected to describe the highly non-linear regime $\xia \gg 1$. To
obtain the statistics of $P(N)$ in cells of radius $R$, volume $V$, it
is convenient to define the quantities:
\beq
S_p = \frac{\xia_p}{\xia^{\; p-1}} \hspace{0.3cm} \mbox{with}
\hspace{0.3cm}
\xia_p =   \int_V \frac{d^3r_1 ... d^3r_p}{V^p} \;
\xi_p ({\bf r}_1,...,{\bf r}_p).
\label{Sp}
\eeq
Then, 
\beq
P(N) = \frac{1}{N_c \; \xia} \; h(x)  \hspace{0.4cm} \mbox{with} \hspace{0.4cm} x = \frac{N}{N_c} = \frac{1+\Delta}{\xia}  \label{PNhx}
\eeq
in the regime where
\beq
N \gg N_v \hspace{1cm} \mbox{and} \hspace{1cm} N_c \gg N_v .
\label {valhx}
\eeq
Here we defined
\beq
N_c = \Na \; \xia , \hspace{0.3cm}
N_v = \frac{\Na }{ \xia^{\;\omega/(1-\omega)}},
\hspace{0.3cm} 
\Na = \na \; V,
\hspace{0.3cm} 
1+\Delta = \frac{N}{\Na }
\eeq
where $\na$ is the average number density in the simulation.
It is also required that the continuous limit $N \gg 1$
is reached in the relevant sampling of the simulation. The two conditions (\ref{valhx}) insure that i) the counts are relevant to mass condensations and not to ``voids'' and ii) one is in the fully non-linear regime where $\xia$ is large so that the non-linear scaling represented by $h(x)$ can develop. The scaling (\ref{PNhx}) implies that
\beq
p \geq 1 \; : \;\; S_p = \int_0^{\infty}  x^p h(x) \; dx
 \hspace{0.3cm}
 , \hspace{0.3cm}  S_1=S_2=1 .  \label{Sphx}
\eeq
From very general considerations one expects that
\beq
\left\{ \begin{array}{rl} x \ll 1 \; : & {\displaystyle  h(x) \sim
\frac{a(1-\omega)}{\Gamma(\omega)} \; x^{\omega-2} } \\ \\  x \gg 1 \;
: &
{\displaystyle h(x) \sim a_s \; x^{\omega_s-1} \; e^{-x/x_s} }
\end{array}
\right.
\label{has}
\eeq
where $x_s$ is a constant, typically of order $10$. Thus, $P(N)$ shows a power-law behaviour in the range $N_v \ll N \ll N_c$ and an exponential cutoff for $N \gg N_c$.

If the scaling laws (\ref{scal1}) apply, then the ratios $S_p$ and the
function $h(x)$ are independent of scale and time. This means that
once $h(x)$ is given (e.g. from measures performed at a certain
scale and time) one only needs to know the evolution of the two-point
correlation function $\xia$ to be able to construct the whole
statistics of $P(N)$ at any scale and time in the highly non-linear
regime. Since $\xia$ obeys (\ref{scal1}), its
evolution is known in the non-linear regime once its normalization is
measured for one scale and time.

\subsection{The two-point correlation function}
\label{The two-point correlation function}

\subsubsection{Fluctuations in a cell}
\label{Fluctuations in a cell}

As we described in the previous section, in order to test the scaling
model in the domain $N \gg N_v$ we first need the evolution of the
two-point correlation function $\xia$.
Moreover, $\xia$ presents a strong interest in itself since it gives a
measure of the amplitude of the density fluctuations and it provides a
first check of the stable-clustering assumption. Indeed, if the latter
holds the slope $\gam$ of $\xia(R)$ must be given by (\ref{gam1}).

\begin{figure}

\psfig{figure=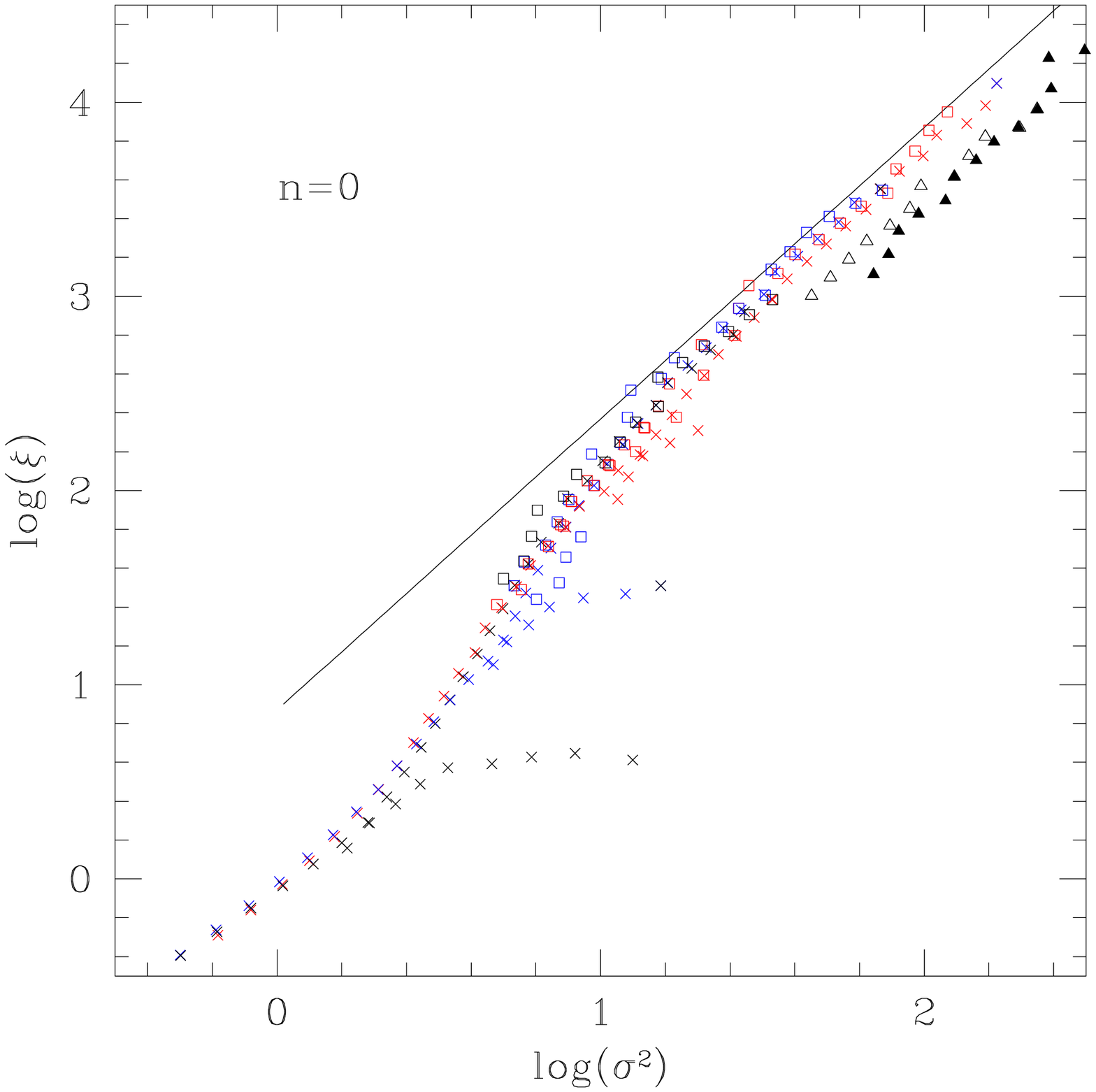,width=8cm,height=6cm}
\psfig{figure=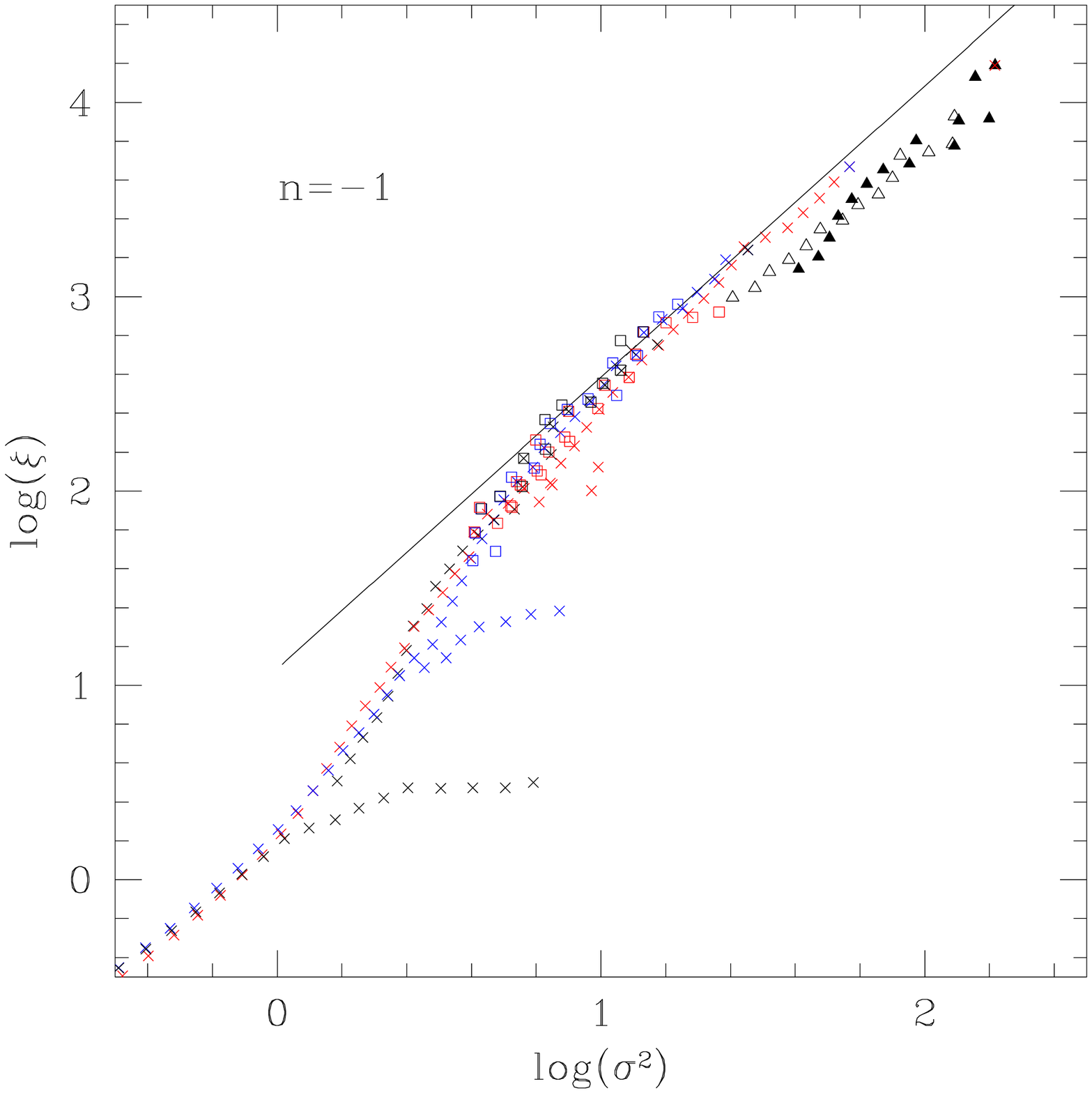,width=8cm,height=6cm}
\psfig{figure=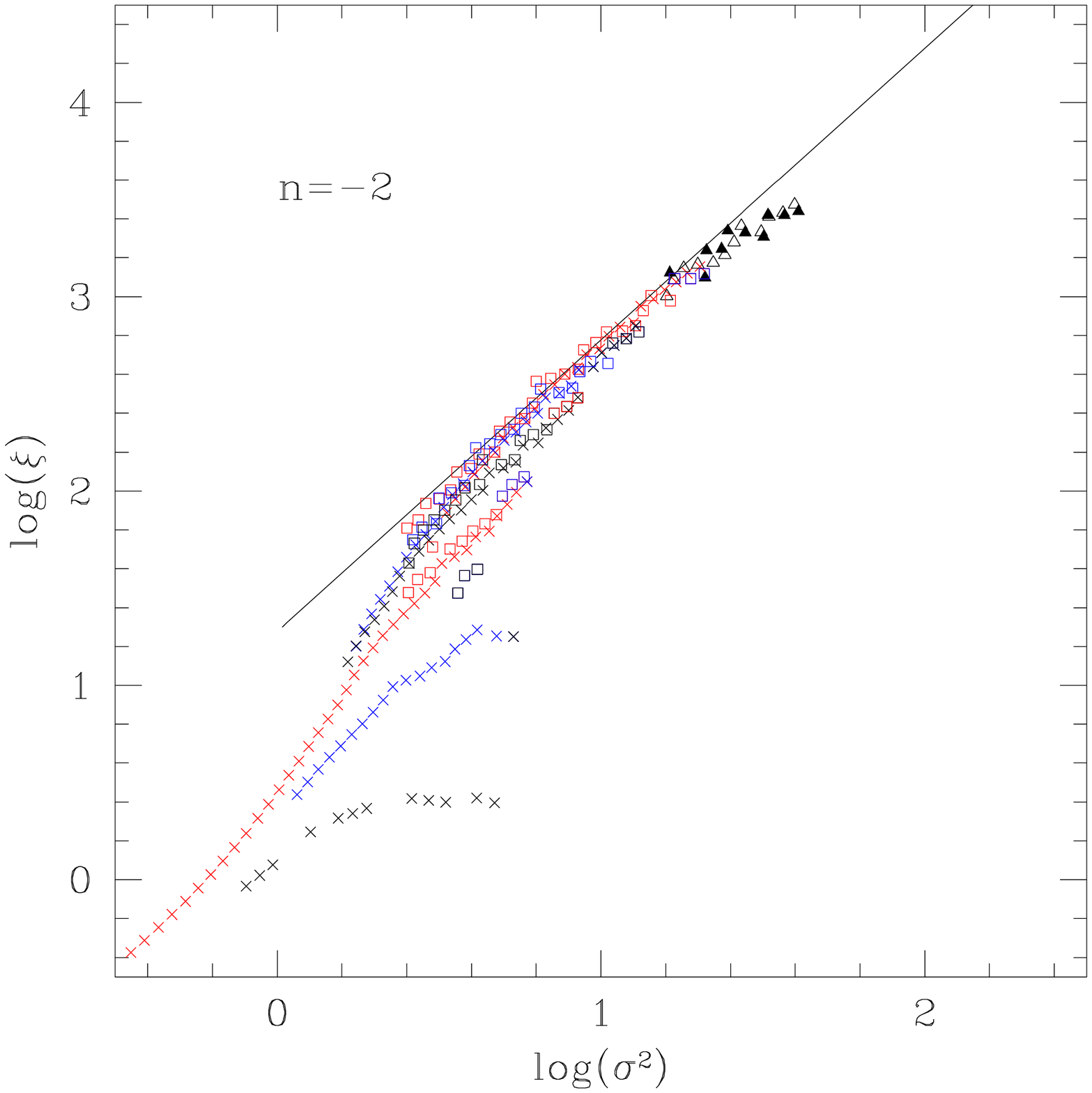,width=8cm,height=6cm}

%{\epsfxsize=8 cm \epsfysize=5.4 cm \epsfbox{figXibtn0.ps} }
%{\epsfxsize=8 cm \epsfysize=5.4 cm \epsfbox{figXibtnm1.ps} }
%{\epsfxsize=8 cm \epsfysize=5.4 cm \epsfbox{figXibtnm2.ps} }

\caption{The two-point correlation function $\xia(R)$ as a function of
$\sigma^2(R_L)$ for the power-spectra $n=0, -1$ and $-2$. The solid
line is the asymptotic behaviour (\ref{Fasym}) for large $\xia$, in
the stable clustering regime, with normalization $\alpha$ from
Tab.1. The crosses are numerical values obtained from counts-in-cells,
while the squares are the estimates of $\xia$ provided by measures of
$\xit$ from neighbour counts, see main text and
(\ref{xitxi}). Different shades of grey correspond to different
comoving scales (0.2, 0.5, 1, 2, 4, 8 and 16 Mpc). The filled (resp. open) triangles show the results for 0.2 Mpc from counts-in-cells (resp. neighbour counts). The larger scales
(8 and 16 Mpc) saturate at too low a value of $\xia$, reflecting the
finite size of the sample. The 0.2 Mpc scale presents
deviations from the scaling in the highly non-linear regime (all curves should
exactly
 superpose for our power-law initial conditions), with a lack of
power due to the softening-over a 0.2 Mpc radius- of the gravitational interaction.}

\label{figXisig}

\end{figure}

To get $\xia(R)$ we simply count the number of particles enclosed in
each of $300^3$ spheres of radius $R$ set on a grid and we measure
$\lag N \rag$ and $\lag N^2 \rag$, where $\lag \rag$ denotes an average over all trials. Then we use the relation:
\beq
\xia = \frac{\lag N^2 \rag}{\lag N \rag^2} - 1 - \frac{1}{\lag N \rag}  \label{xiN2}
\eeq
As argued by VS, following the ideas of
Hamilton et al.(1991)
from a
Lagrangian point of view where one follows the evolution of matter
elements,
one expects $\xia(R,a)$ to be closely related to the linear
correlation function $\sigma^2(R_L,a)$ evaluated at a different scale
$R_L$ but at the same time:
\beq
\left\{ \begin{array}{rcl} \xia(R,a) & = & {\displaystyle F
\left[ \sigma^2(R_L,a) \right] } \\  \\ R_L^3 & = & {\displaystyle
 \left[ 1+\xia(R,a) \right] \; R^3 }  \end{array} \right.
  \label{Xisig}
\eeq
In fact, since our initial conditions are scale-invariant the relation
(\ref{Xisig}) {\it is exact}: it is then just a rewriting of the scaling solution of Peebles (1980) that holds in this case. 
Its main interest comes from its physical
interpretation which suggests that the $F(\sigma^2)$ obtained for
various $n$
should show a similar behaviour. Of course in the linear regime $\xia
\ll 1$ we must recover $\xia \simeq \sigma^2$ while in the highly
non-linear regime if clustering is stable we have $\xia(R,a) \propto
a^3$. This gives the asymptotic behaviour of the function
$F(\sigma^2)$:
\beq
\left\{  \begin{array}{rcl}   \sigma \ll 1 & : & F(\sigma^2) = \sigma^2
 \\ \\ \sigma \gg 1 & : &
F(\sigma^2) = \left(\frac{10}{3\alpha}\right)^3 \; \sigma^{\;3}
\end{array} \right.
\label{Fasym}
\eeq
where $\alpha \simeq 1$. As a consequence, we plot the values of
$\xia(R)$ obtained from the simulations through (\ref{xiN2}) as a
function of $\sigma^2(R_L)$ rather than $R$.
 Thus, at a given scale
$R$ we measure $\xia(R)$ and we know $\sigma^2(R)$ from which we
derive $\sigma^2(R_L) = \sigma^2(R) \; (R_L/R)^{-(n+3)} = \sigma^2(R)
\; (1+\xia)^{-(n+3)/3}$. The results are shown in Fig.\ref{figXisig}
for
various $n$.

\begin{table}
\begin{center}
\caption{Normalization parameters for the two-point correlation
functions.
The non-linear exponent $\gam$ is given by (\ref{gam1}). The columns
(J) and (C) correspond to the numerical results obtained by Jain et
al.(1995) and Colombi et al.(1996), see text. Note that the ratio
$\alpha/\tilde{\alpha}$ is predicted by the stable-clustering ansatz,
see (\ref{alpalpt}).}
\begin{tabular}{ccccccccc}\hline

$n$ & $\gam$ & $\alpha/\tilde{\alpha}$ & $\alpha$ & $\alpha$ (J) &
$\alpha$ (C) & $\tilde{\alpha}$ & $\tilde{\alpha}$ (J) &
$\tilde{\alpha}$ (C) \\
\hline\hline
\\

0 & 1.8 & 1.10 & 1.71 & 1.43 & 1.28 & 1.55 & 1.30 & 1.16 \\

-1 & 1.5 & 0.95 & 1.45 & 1.18 & 1.33 & 1.52 & 1.24 & 1.40 \\

-2 & 1 & 0.96 & 1.25 & 1.09 & 1.25 & 1.30 & 1.13 & 1.30 \\

\end{tabular}
\end{center}
\label{table1}
\end{table}

Since the initial conditions of the simulation are scale-invariant
($P(k)$ is a power-law) all curves should superpose as long as
numerical effects (finite box size, softening length and resolution
scale) are small. We can check that this is indeed the case. Moreover,
in the highly non-linear regime we recover the asymptotic behaviour
given by (\ref{Fasym}) which is shown by the solid line. The
normalization parameter $\alpha$ is displayed in Tab.1 for the three
power-spectra. However, we can note that for larger scales the
two-point correlation function ``saturates'' to a value lower than its
asymptotic limit. This problem increases for smaller $n$ and is due to
numerical effects since as we explained above all curves should
superpose. The dependence on scale and on $n$ of this effect shows that it is produced by the lack of power in the simulation at
large scales. Indeed, the actual initial power-spectrum $P(k)$ is a
power-law only over a limited range of $k$ and the influence of the
numerical cutoff at $k_{min}$, unduely supressing some power, becomes
more important as one considers larger scales (hence wavenumbers
closer to $k_{min}=2\pi/L$) and lower $n$ (which increases the
contribution of small $k$). Similarly, there is a high frequency
cut-off. The correlation function $\xi$ starts to deviate from the
exact scaling
imposed by our power-law initial conditions when the
sampling of the numerical output is done at the $0.2$ Mpc comoving
scale and below. This is very distinctly seen in Fig.\ref{figXisig}
for the larger values of the correlation function. This is due to the
softening of the interaction, which for all the samples we use is
done over a radius of
$0.2$ Mpc (comoving). Again, the result is a lack of power, due to
this softening. Obviously, the deficit is larger for $n = 0$ than for
$n=-2$, the former case, with more small-scale power, being more
sensitive to this effect.

Note that the Nyquist frequency and particle discreteness play a minor
role if any at all since we only consider the cases where the number
of points in the simulation is large enough.

\subsubsection{Counts of neighbours}
\label{Counts of neighbours}

Although the average $\xia$ defined by (\ref{Sp}) appears naturally
when one considers counts in cells many authors studied the quantity
$\xit$ given by:
\beq
\xit(R) = \frac{3}{R^3} \; \int_0^{R} \xi(r) \; r^2 \; dr
\eeq
which is relevant for the counts of neighbours. Indeed, the mean
number of neighbours located within a radius $R$ from a given particle
is:
\beq
\lag N_{nb} \rag \;\; = \;\; \lag N \rag \; \left( 1 + \xit(R) \right)  \label{Nnb}
\eeq

\begin{figure}

\centering \psfig{figure=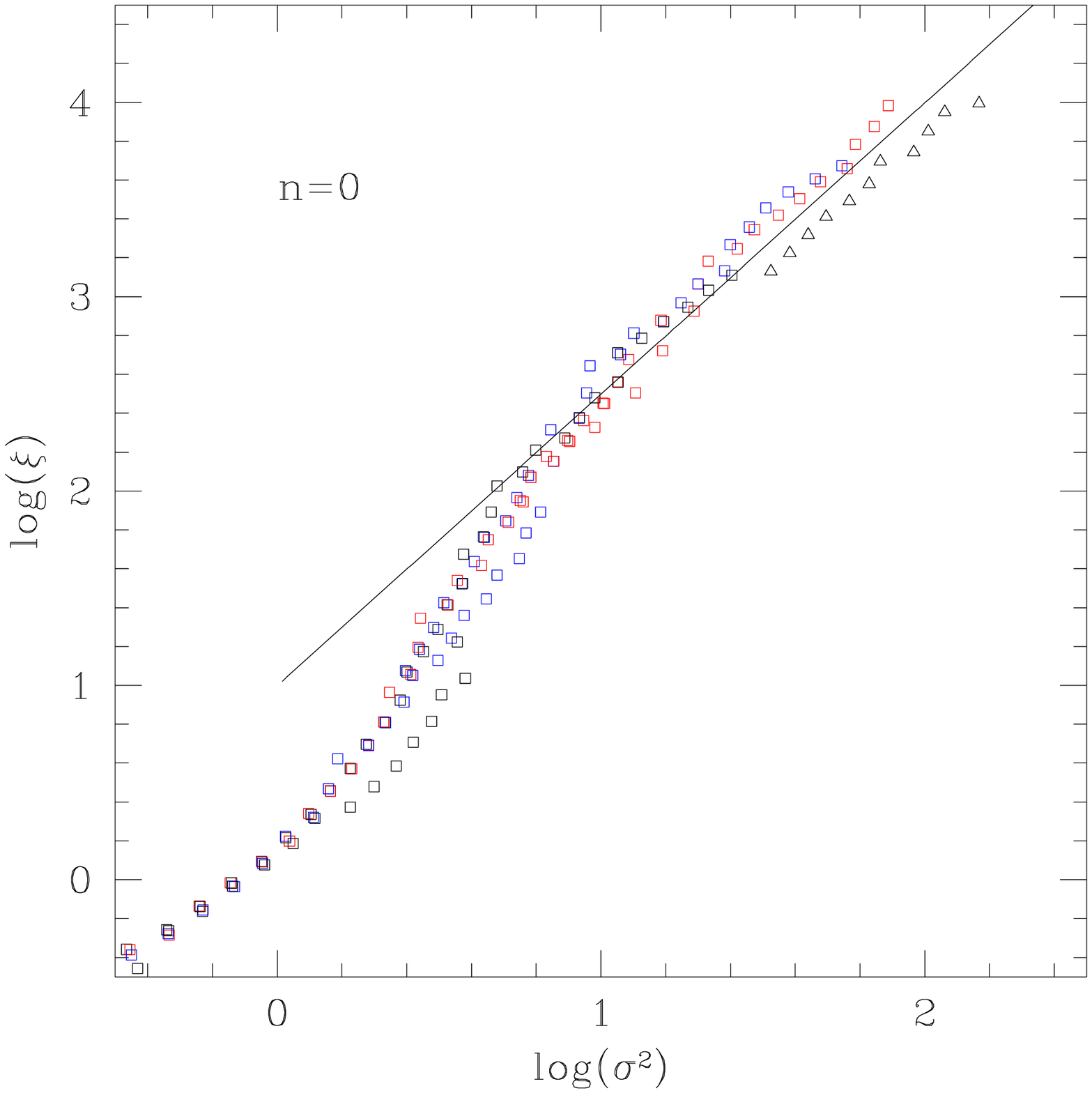,width=8cm,height=6cm}
\centering \psfig{figure=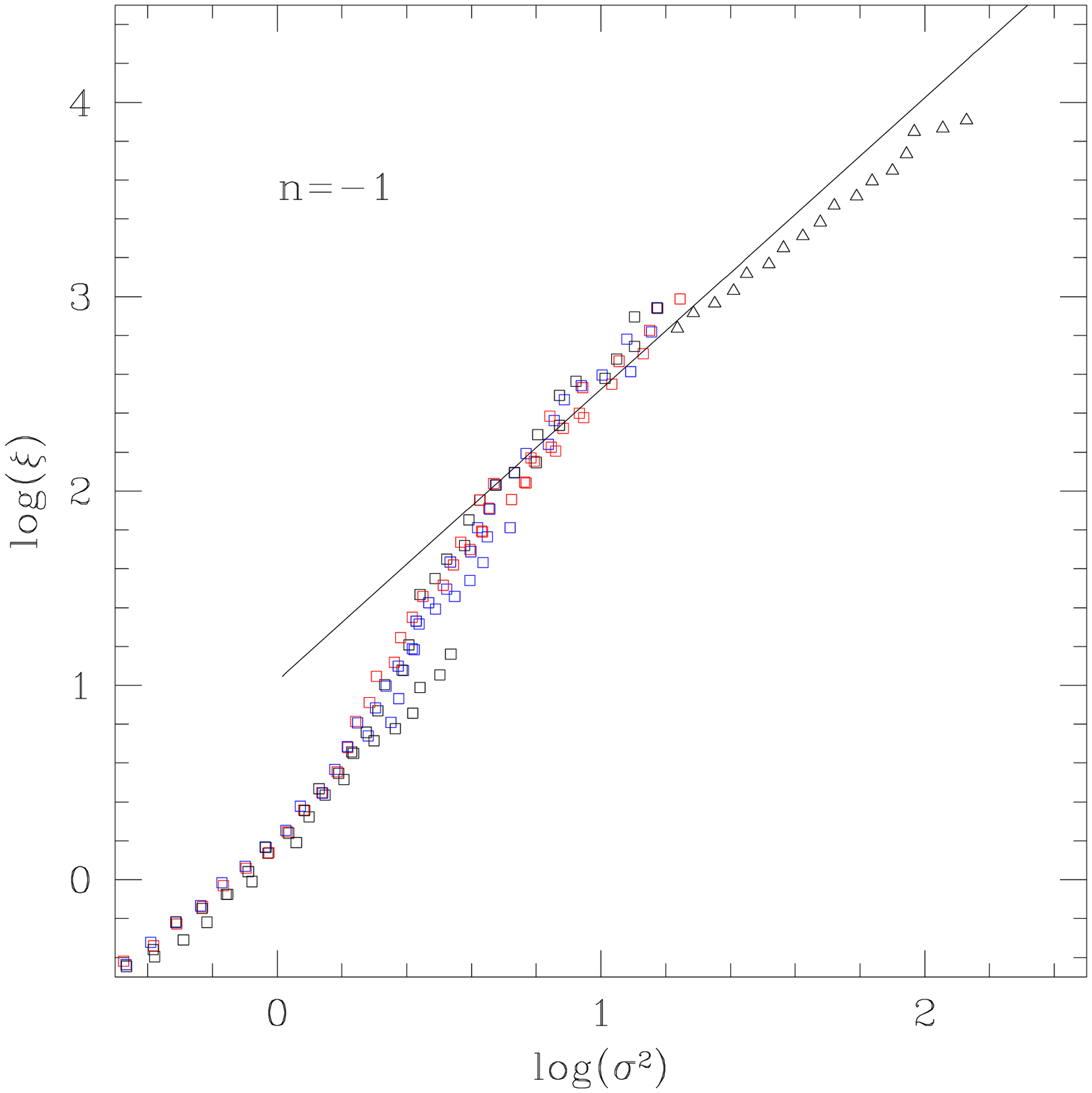,width=8cm,height=6cm}
\centering \psfig{figure=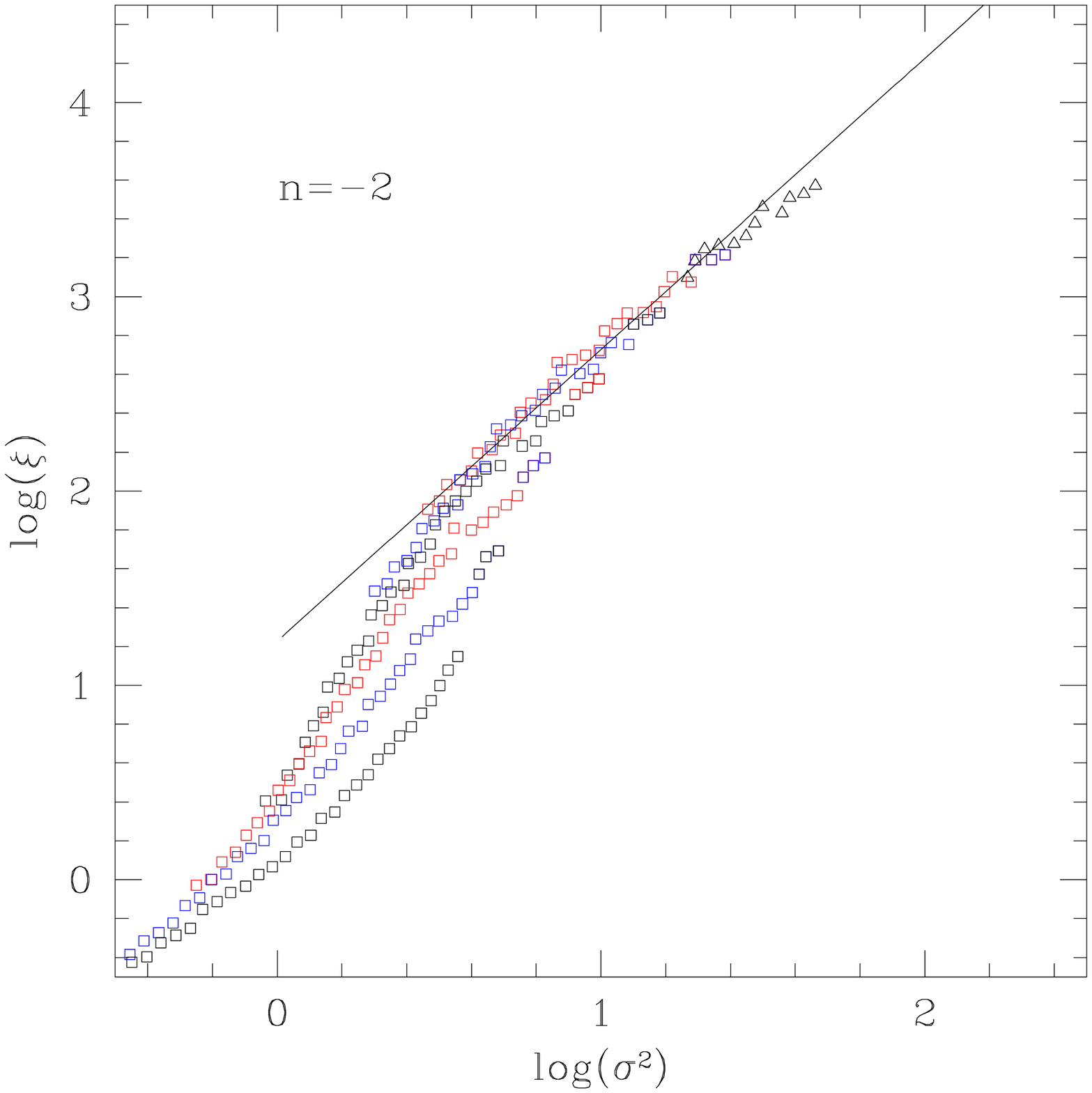,width=8cm,height=6cm}

%{\epsfxsize=8 cm \epsfysize=5.5 cm \epsfbox{figXitsigtn0.ps} }
%{\epsfxsize=8 cm \epsfysize=5.5 cm \epsfbox{figXitsigtnm1.ps} }
%{\epsfxsize=8 cm \epsfysize=5.5 cm \epsfbox{figXitsigtnm2.ps} }

\caption{The two-point correlation function $\xit(R)$ (obtained
from neighbour counts) as a function of $\sigmat^2(R_L)$ for the
power-spectra $n=0, n=-1$ and $n=-2$. The solid line is the asymptotic
behaviour (\ref{Fasym}) for large $\xit$, with the normalization
parameter $\tilde{\alpha}$ from Tab.1. The triangles correspond to the comoving scale 0.2 Mpc. See Fig.\ref{figXisig} and the
text for comments.}
\label{figXitsigt}

\end{figure}

For a power-law power-spectrum $\xit$ depends on its linear equivalent $\sigmat^2$ in a fashion similar to (\ref{Xisig}) with a new function $\Ft$ which has the same asymptotic behaviour as (\ref{Fasym}) but with a
slightly different normalization $\tilde{\alpha}$ (Hamilton et al.1991; Peebles 1980; see also Padmanabhan 1996). We measure $\xit$
in the simulation using (\ref{Nnb}): we count the mean number of
neighbours of particles in the box which gives $<N_{nb}>$ while the
total number of points in the simulation provides $<N>$. The result is
shown in Fig.\ref{figXitsigt}. We indeed recover the
behaviour predicted by (\ref{Fasym}). This confirms the previous
results of Fig.\ref{figXisig} since both $\xia$ and $\xit$ are
measures of $\xi$. This shows that in the non-linear regime the
stable-clustering ansatz holds for $\xi$ which obeys the scaling-law
(\ref{scal1}).

Moreover, the measure of $\xit$ allows us to get a second estimate of
$\xia$ in the non-linear regime. Indeed, in the domain where $\xia \gg
1$ the two-point correlation function is a power-law with a slope
$\gam$ given by (\ref{gam1}) as shown by Fig.\ref{figXisig} and
Fig.\ref{figXitsigt}. In this case, one can show (Peebles 1980) that:
\beq
\xit(R) = \beta \; \xia(R) \hspace{0.3cm} \mbox{with} \hspace{0.3cm}
\beta = (1-\gam/4) \; (1-\gam/6) \; 2^{\gam}
\label{xitxi}
\eeq
Thus, we can obtain $\xia(R)$ from the measure of $\xit(R)$. These
values are shown by squares in Fig.\ref{figXisig}. We can check that
they agree with the previous calculations for $\xia$ based on the
counts in cells. Note that we only show in Fig.\ref{figXisig} the
values of $\xia$ obtained from $\xit$, through (\ref{xitxi}), which
are in the highly non-linear regime $\xia \gg 1$ where the two-point
correlation function is a power-law so that (\ref{xitxi}) applies.
Note that in this case (compare Fig.\ref{figXisig} with
Fig.\ref{figXitsigt}), the lack of power at the larger scales is much
less pronounced: the count of neighbours already focuses on the dense
regions and the statistics of the poorly represented events at large
scale is improved. Indeed, to measure $\xit$ one considers cells which
are centered on particles, hence this statistical tool follows the
evolution of gravitational clustering as it automatically probes more
closely denser regions, while to get $\xia$ the center of the cells is
set at random in the box, so that in the highly non-linear regime most
cells are within voids. Thus at least part of the deviations seen at
these scales are due to the statistical extraction of the information.

We present in Tab.1 the values of the normalization parameters
$\alpha$ and $\tilde{\alpha}$ obtained from our simulation, as well as
from Jain et al.(1995) and Colombi et al.(1996). For the former we
calculate $\alpha$ from their value of $\tilde{\alpha}$ while for the
latter we obtain $\tilde{\alpha}$ from $\alpha$. Indeed, from
(\ref{xitxi}) we get:
\beq
\alpha = \tilde{\alpha} \; \beta^{1/3} \; \beta_L^{-1/2}
\label{alpalpt}
\eeq
where $\beta$ (resp. $\beta_L$) is given by (\ref{xitxi}) with
$\gam=3 (3+n)/(5+n)$ (resp. $\gam=(3+n)$), corresponding to the
non-linear (resp. linear) regime. Although all numerical simulations
agree with the stable-clustering ansatz: (\ref{gam1}), (\ref{Xisig})
and (\ref{Fasym}), the normalization of $\xia$ in the non-linear
regime varies up to a factor 2 (note that it scales as the cube of
$\alpha$). Thus, there is still some inaccuracy in the numerical
values of $\xia$. Moreover, one may expect the higher-order
correlation functions (hence the parameters $S_p$) to bear at least
similar uncertainties. The discrepancies between the various estimates
of $\alpha$ may be due to the effects of finite volume and particle
number: in particular the initial power-spectrum is not a power-law
over an infinite range of scales (there are an upper and a lower
cutoff) and one should average $\xia$ over many realizations of the
initial gaussian field. However, the fact that curves obtained from
different scales superpose in Fig.\ref{figXisig} and
Fig.\ref{figXitsigt} and that our measures of $\xit$ are consistent
with $\xia$ show that we can reasonably rely on our results.
Except in the extreme cases where the finiteness of the sample (at large scales) or the smoothing of the force (at small scales) become important, 
{\it these results are consistent with the idea that the stable clustering regime is reached} (see however Padmanabhan et al.1996 for an alternative view).

\subsubsection{Analytical fit to the fluctuations in a cell}
\label{Analytical fit to the fluctuations in a cell}

\begin{figure}

\centering \psfig{figure=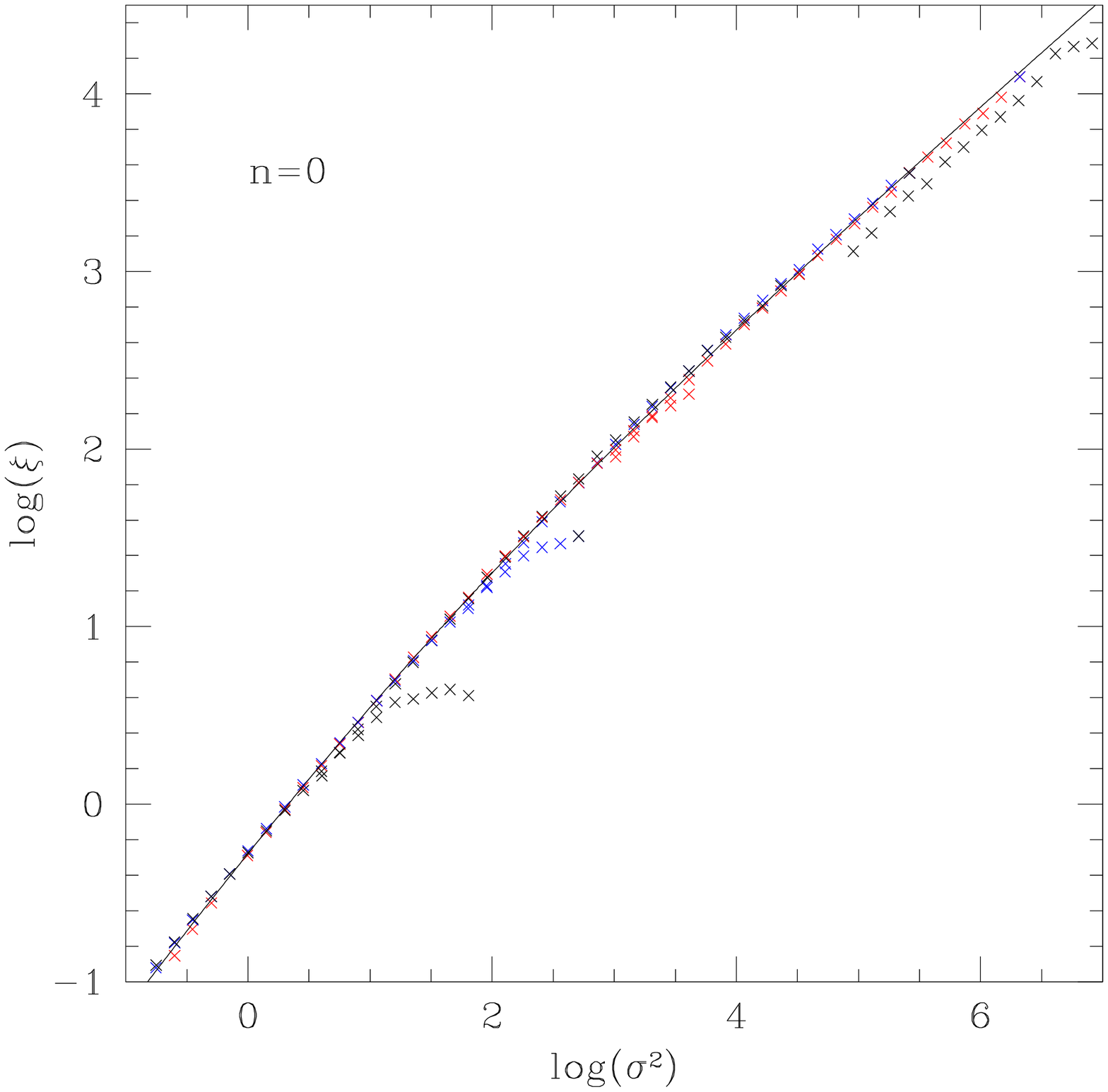,width=8cm,height=6cm}
\centering \psfig{figure=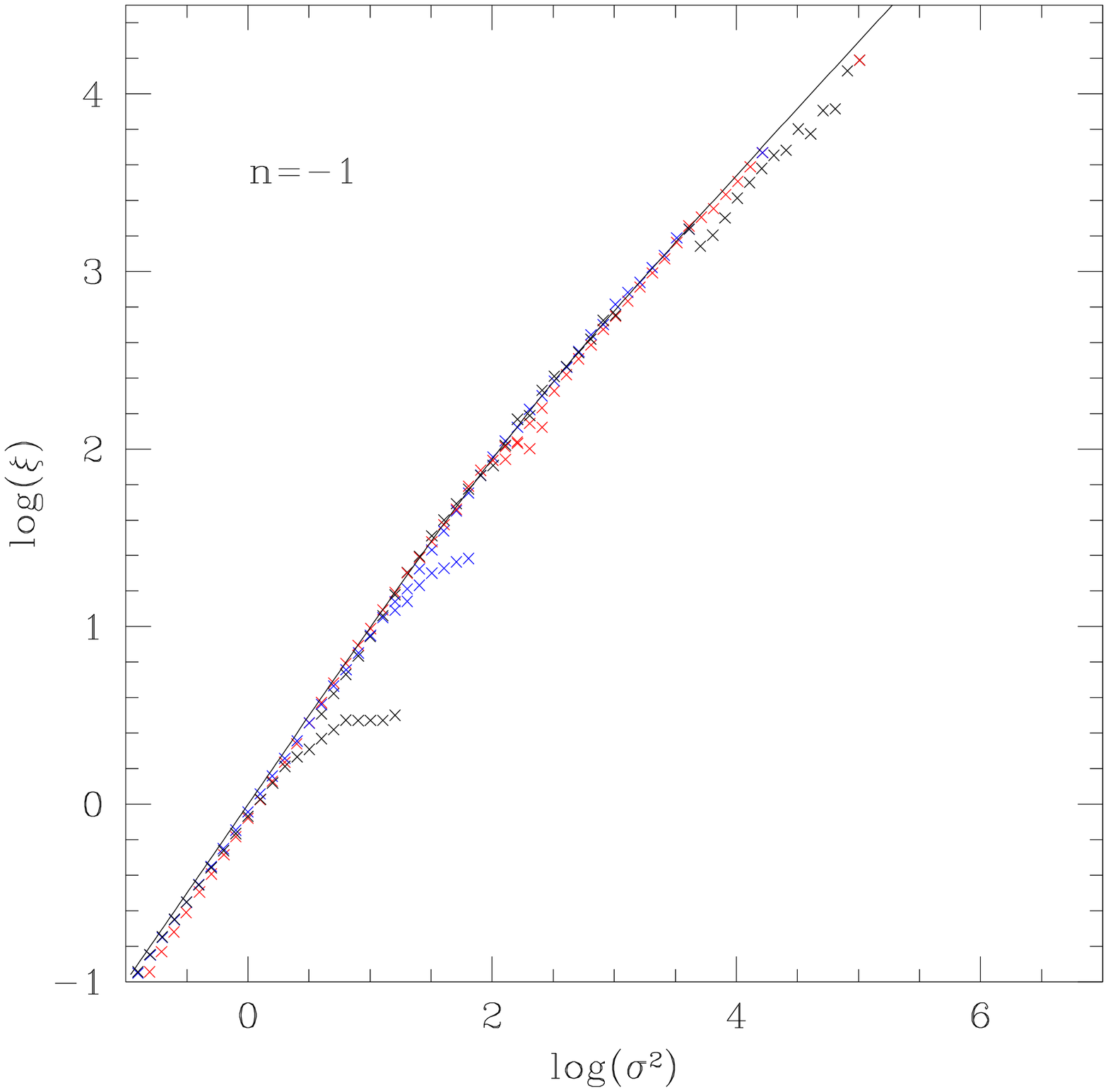,width=8cm,height=6cm}
\centering \psfig{figure=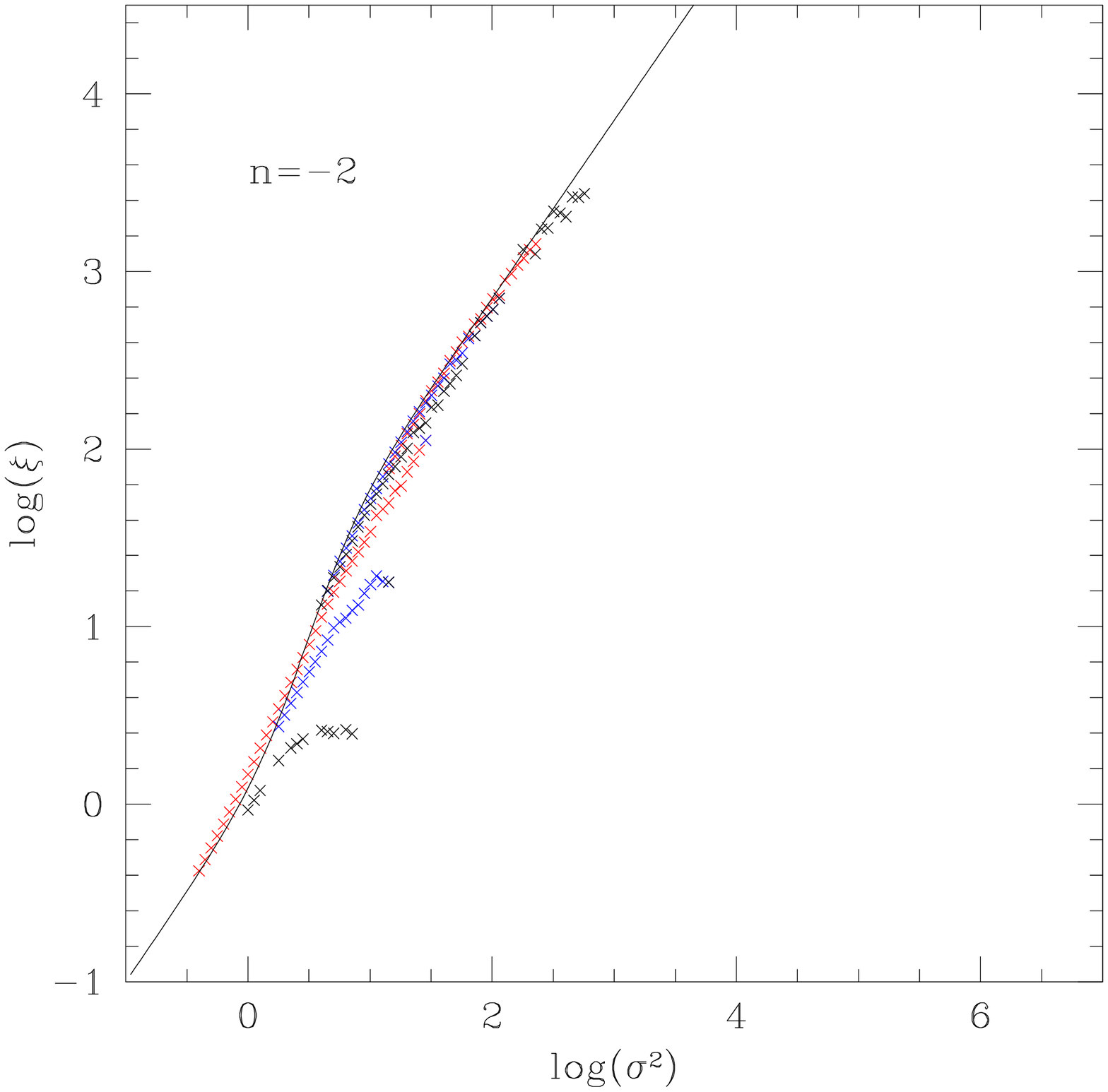,width=8cm,height=6cm}

%{\epsfxsize=8 cm \epsfysize=5.5 cm \epsfbox{figXisigRn0.ps} }
%{\epsfxsize=8 cm \epsfysize=5.5 cm \epsfbox{figXisigRnm1.ps} }
%{\epsfxsize=8 cm \epsfysize=5.5 cm \epsfbox{figXisigRnm2.ps} }

\caption{The two-point correlation function $\xia(R)$ as a function of
$\sigma^2(R)$ (hence taken at the same spatial scale). The crosses are
numerical estimates from counts in cells in the simulations while the
solid lines are the analytical fits (\ref{fitXi}). See
Fig.\ref{figXisig} and text for comments.}
\label{figXisigR}

\end{figure}

We show in Fig.\ref{figXisigR} the relation $\sigma^2(R)
\leftrightarrow \xia(R)$. This corresponds to an Eulerian point of
view as opposed to the Lagrangian point of view displayed in
Fig.\ref{figXisig}, since we now consider the values of the linear and
non-linear correlation functions taken at the same spatial scale (and
not at the same ``mass scale'').
We can note that the curves we obtain are quite smooth when displayed
in this diagram and they do not show a sharp ``step-profile'' as was
the case in Fig.\ref{figXisig} and Fig.\ref{figXitsigt}. The solid
lines are analytical fits to the numerical results which we shall need
below (Sect.\ref{Mass functions}) to get the mass functions within the
framework of the scaling approach. Thus we use the approximations:
\beq
\left\{  \begin{array}{rl}  n = 0, 1 : & {\displaystyle \xia(R) \simeq
\left[ (\sigma^2)^{-a} + (\xia_1)^{-a} \right]^{-1/a} } \\  \\
\mbox{with} & n=0 :  a=0.7 \hspace{0.3cm} ; \hspace{0.3cm} n=-1 :  a=4
 \\  \\ n = -2 : & {\displaystyle
\xia(R) \simeq \left[ \sigma^2 + (\sigma^2/5)^2 \xia_1 \right]
/ \left[ 1 + (\sigma^2/5)^2 \right] } \end{array} \right.
 \label{fitXi}
\eeq
where $\xia_1(R) = ( \frac{10}{3\alpha} \; \sigma(R) )^{6/(5+n)}$. Of
course, these fits are built so as to be consistent with  (\ref{Xisig})
and
(\ref{Fasym}). These relations  define implicitly the
functions $F(\sigma^2)$. They will be used
instead of the correlation function $\xia$ actually measured in the
simulation (hence they will not be a source of unwanted deviations from
self-similarity) when we determine the scaling properties of the
multiplicity functions (Sect.\ref{Mass functions}).

\subsection{Counts in cells statistics}
\label{Counts in cells statistics}

\begin{figure}

\centering \psfig{figure=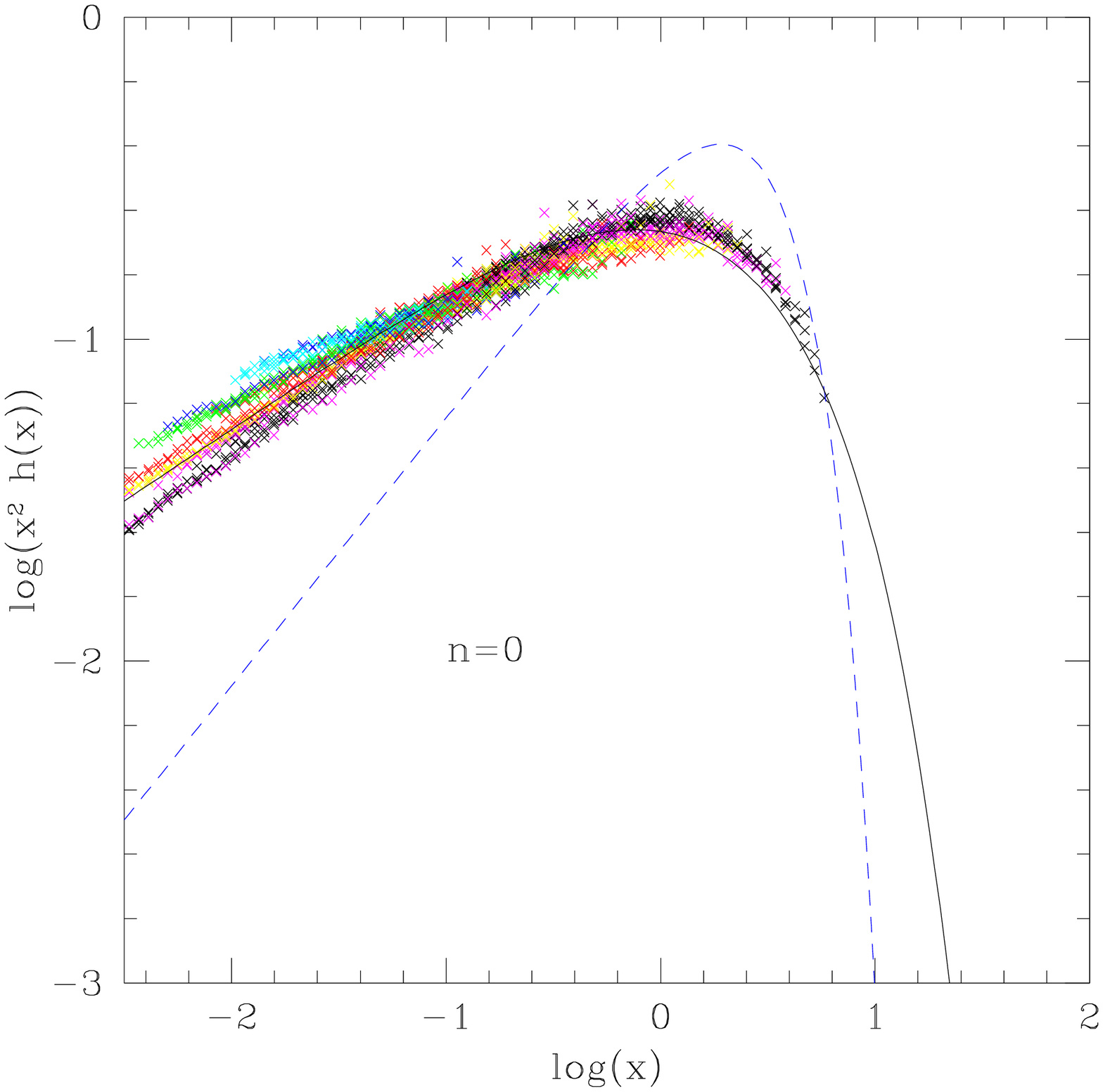,width=8cm,height=6cm}
\centering \psfig{figure=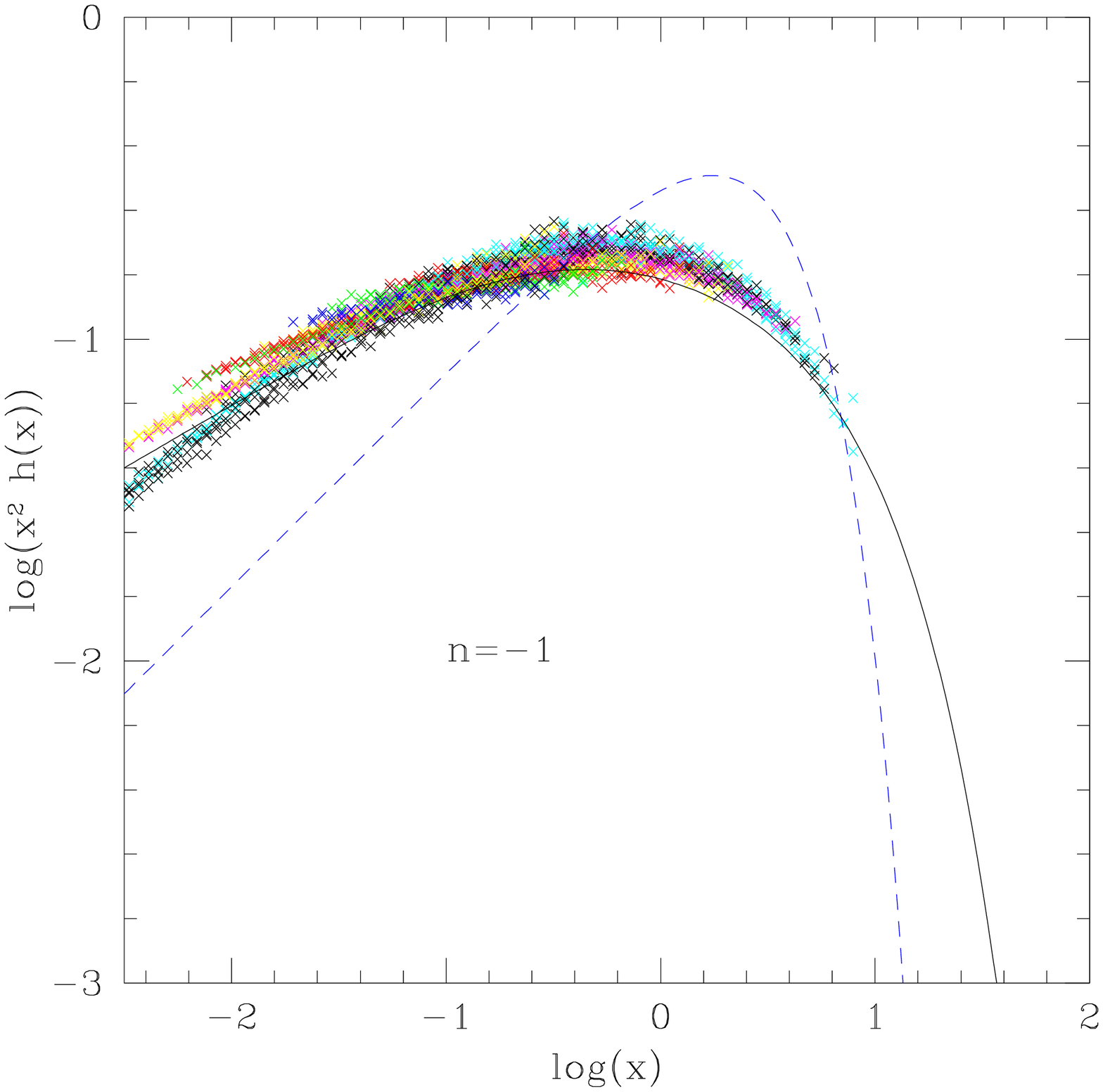,width=8cm,height=6cm}
\centering \psfig{figure=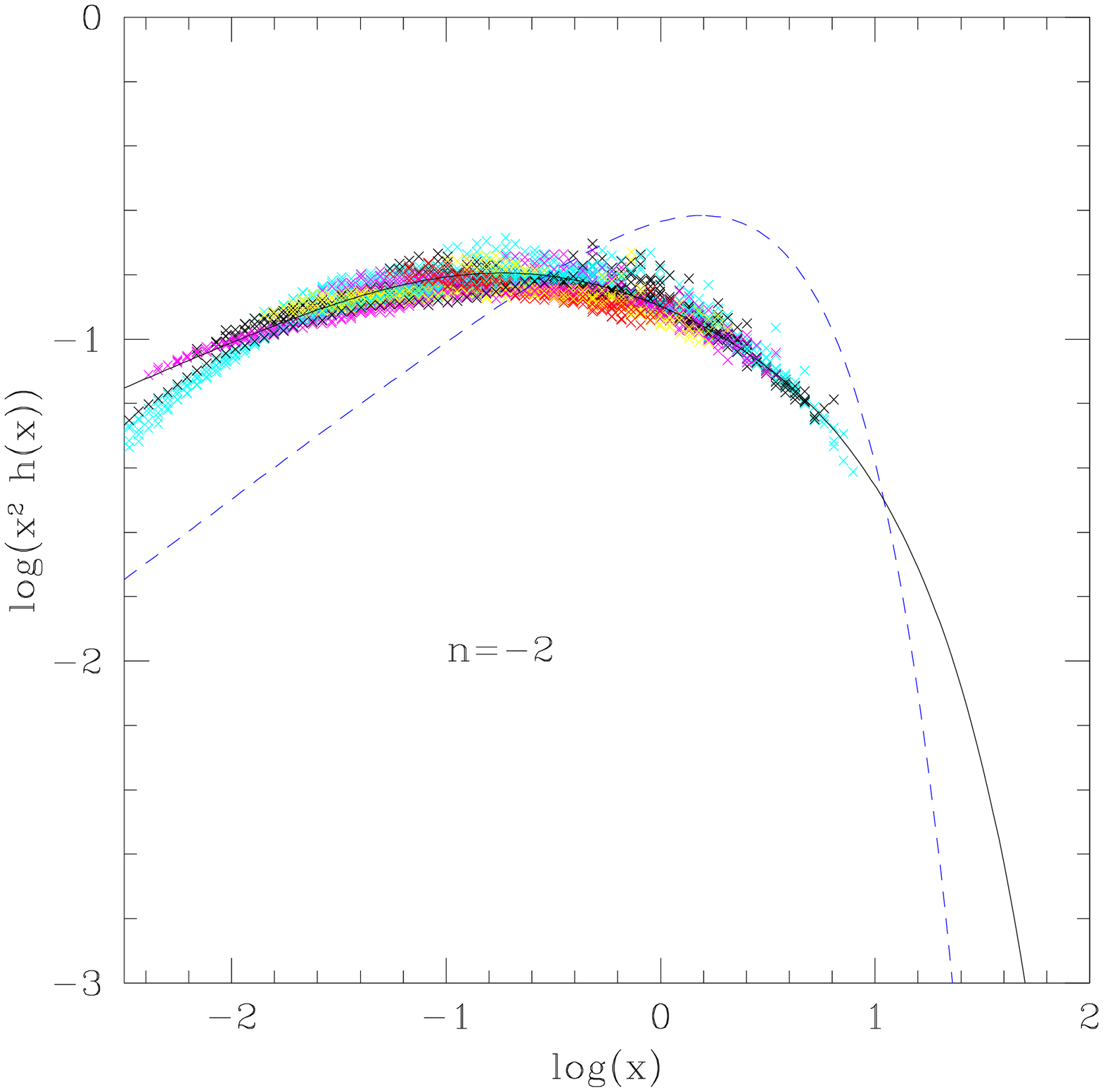,width=8cm,height=6cm}

%{\epsfxsize=8 cm \epsfysize=5.5 cm \epsfbox{figPNn0.ps} }
%{\epsfxsize=8 cm \epsfysize=5.5 cm \epsfbox{figPNnm1.ps} }
%{\epsfxsize=8 cm \epsfysize=5.5 cm \epsfbox{figPNnm2.ps} }

\caption{The probability distribution $P(N)$ for counts in cells. We
display the quantity $N^2/\Na \; P(N) = x^2 h(x)$ as a function of
$x=N/N_c$. The solid curves are the analytical fits given by
(\ref{fithx}) and Table~2 while the dashed lines show the ``PSs'' scaling
function (\ref{hPSs}) obtained using the Press-Schechter approach
in the regime where the correlation function has reached stable
clustering. Different shades of grey correspond to different values of
$\xia$, ranging from 100 up to the highest value shown in
Fig.\ref{figXisig}. Only the scales (from 0.5 to 4 Mpc) where the
correlation function is seen (Figs.\ref{figXisig}, \ref{figXitsigt},
\ref{figXisigR}) to obey the scaling required by the exact
self-similarity are considered here and in the following.}
\label{figPN}

\end{figure}

By counting the number of particles enclosed in each of $300^3$
spheres set on a grid, as we did to evaluate $\xia$, we can also
obtain the statistics of the counts in cells. This allows us to
compare the predictions (\ref{PNhx}) of the scaling model to our
numerical results for $P(N)$. Note that we obtain simultaneously the
value of $\xia$ in the simulation, at the scale of interest, which we
use in (\ref{PNhx}). We display the results in Fig.\ref{figPN} as a
function of $x$: from $N$ and $P(N)$ obtained by these counts in the
simulation we define:
\beq
x = \frac{N}{N_c} \hspace{0.3cm} \mbox{and} \hspace{0.3cm}
x^2 h(x) = \frac{N^2}{\Na} \; P(N)  \label{PNx}
\eeq
If the scaling-laws (\ref{scal1}) hold the quantity $h(x)$ used above
must be the scaling function defined in (\ref{PNhx}). Then, all curves
obtained for different sizes and times should superpose. Note that in
our case where the initial power-spectrum is a power-law all curves
characterized by the same $\xia$ (or equivalently by the same
$\sigma^2$) must coincide. Thus the scaling-laws (\ref{scal1}) merely
imply in addition that curves measured for different $\xia$ should
also superpose, when shown in a diagram $x \leftrightarrow x^2h(x)$ as
defined by (\ref{PNx}).

We can see in Fig.\ref{figPN} that the various curves indeed
superpose, for all three power-spectra. Note that in this figure we
only show the parts of $P(N)$ which should scale as predicted by
(\ref{PNhx}): we only display points which satisfy the requirements
$\xia>100, N>4, N>4 N_v$ and $N_c>4$. The solid curves are analytical
fits to the data points. We use the functional form introduced by
Bouchet et al.(1991):
\beq
h(x) = \frac{a(1-\omega)}{\Gamma(\omega)} \;
\frac{x^{\omega-2}}{(1+bx)^{c}}
\; \exp(-x/x_s)    \label{fithx}
\eeq
The values of the parameters $a, \omega, b, c$ and $x_s$ are given in
Tab.2. Note that the functions $h(x)$ must also obey the constraints:
\beq
\int_0^{\infty}  x \; h(x) \; dx = \int_0^{\infty}  x^2 \; h(x) \; dx =
1
\label{hnorm}
\eeq
from the general relation (\ref{Sphx}).

The dashed curve in the figures is the function $h(x)$ which one
obtains (see VS) by assuming that the non-linear evolution of initial
density fluctuations is given by the Press-Schechter approximation
in the case where the correlation function has reached the stable
clustering regime (which is indeed the case in the examples we have
considered in Fig.\ref{figPN}). Thus, we obtain what we shall refer to
as the ``PSs'' estimate (``s'' for stable)
\beq
h_{PSs}(x) = \sqrt{\frac{2}{\pi}} \; \frac{5+n}{6\alpha} \;
x^{(n+5)/6-2}
 \; \exp \left( - \frac{x^{(5+n)/3}}{2\alpha^2} \right)
\label{hPSs}
\eeq
the result being multiplied by the standard ``PS factor'' of two. 
For the stable clustering regime, $x < (1+\Delta) / 200$, this expression is identical to the usual Press-Schechter formula (normally used for $\Delta \sim 177$, the usual density contrast of just-virialized halos).
If the above condition on $x$ is not fulfilled, only minor
differences between the two expressions appear however, see Fig.\ref{figPS} below.
We can see in Fig.\ref{figPN} that the function $h_{PSs}(x)$ obtained in this
way does not agree with the numerical results, the true distribution
being much broader as argued by VS (the general trend, however, larger
$\omega$ and sharper cutoff for higher $n$ is correct). This means
either that one cannot follow the evolution of individual matter
elements with using only the spherical model, because there is some
exchange of matter between neighbouring density fluctuations through
mergers and disruptions and the corrections to the spherical dynamics
are not negligible, or that after virialization the density
fluctuations still undergo non-negligible evolution.
The stable-clustering
ansatz on the other hand is seen
 from
Fig.\ref{figXisig} and Fig.\ref{figPN} to be
 a valid approximation in a statistical sense.

\begin{table}
\begin{center}
\caption{Parameters for the scaling function h(x).}
\begin{tabular}{cccccc}\hline

$n$ & $a$ & $\omega$ & $x_s$ & $b$ & $c$ \\
\hline\hline
\\

0 & 1.50 & 0.45 & 4 & 1 & 0.6  \\

-1 & 1.48 & 0.4 & 8 & 3 & 0.6\\

-2 & 1.71 & 0.3 & 13 & 5 & 0.6 \\

\end{tabular}
\end{center}
\label{table2}
\end{table}

The scaling of the count-in-cells has thus been checked over an
unprecedented
range (more than 3 decades in the scaling parameter $x$). Note that we
check here not only the scaling of the counts as a function of the cell
size $R$ but also as a function of time, that is the self-similarity
 -in principle exact-
of the computation results.

\begin{figure}

\centering \psfig{figure=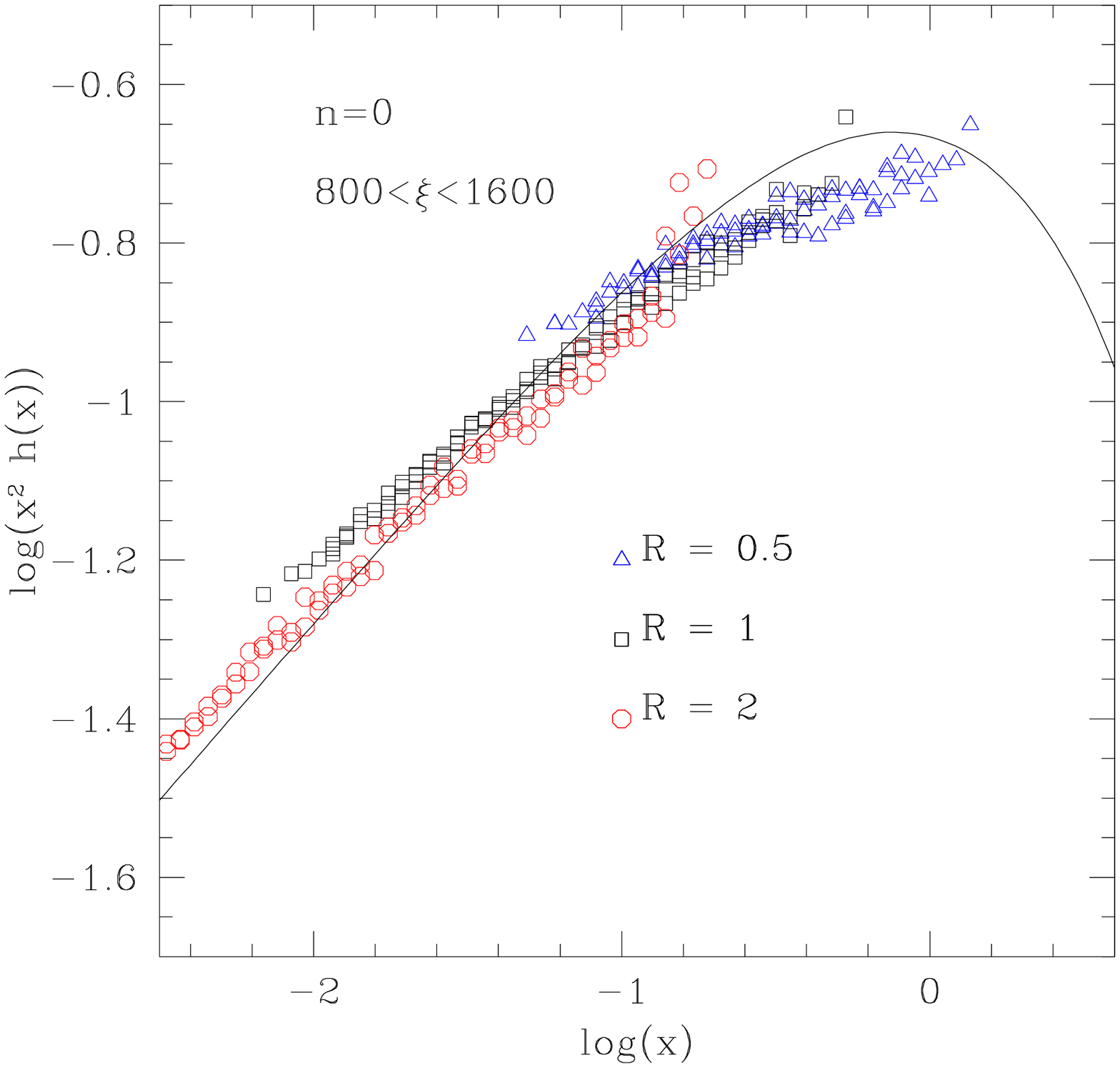,width=8cm,height=6cm}
\centering \psfig{figure=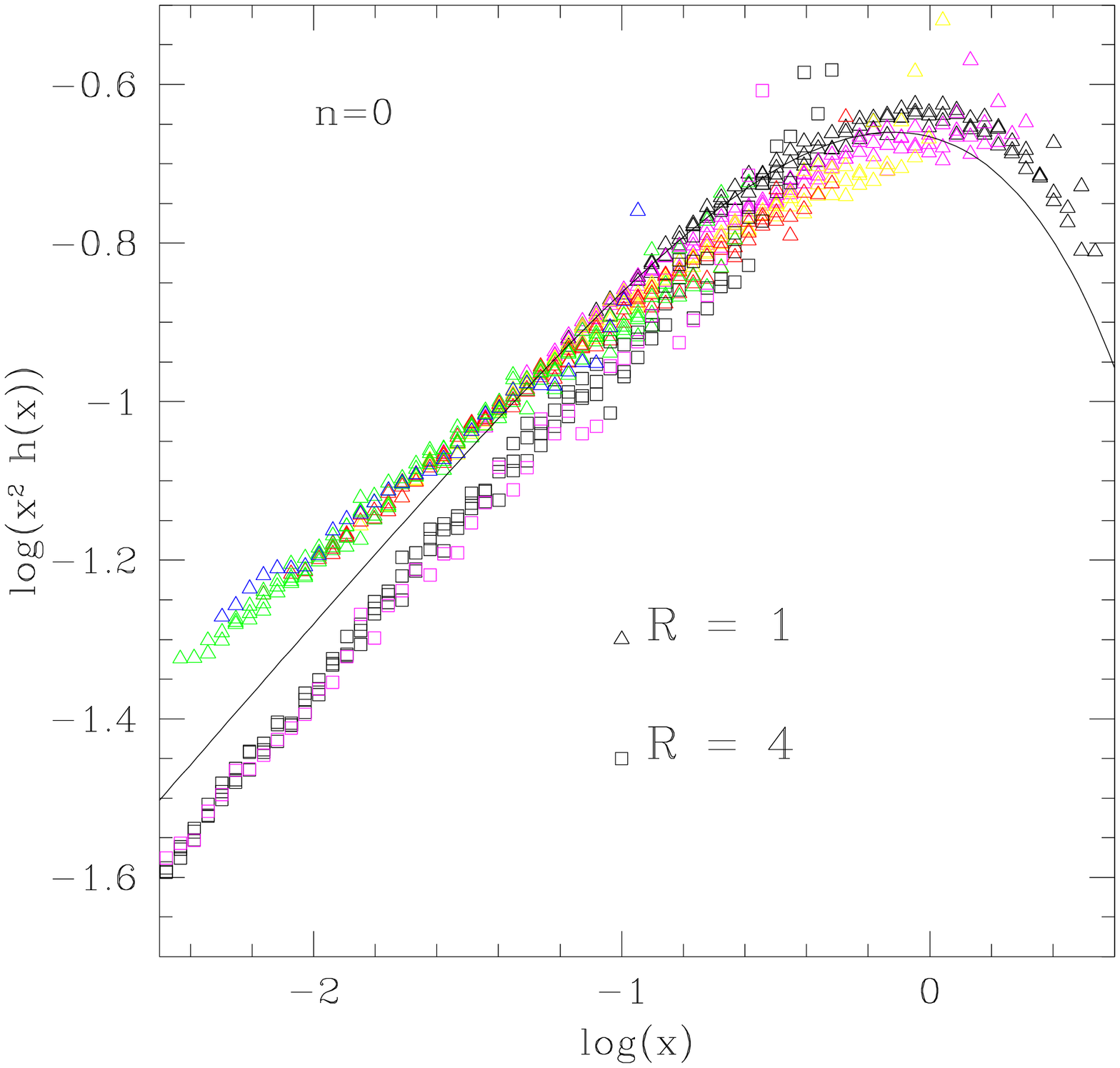,width=8cm,height=6cm}

%{\epsfxsize=8 cm \epsfysize=5.6 cm \epsfbox{figPNtestXin0.ps} }
%{\epsfxsize=8 cm \epsfysize=5.6 cm \epsfbox{figPNtestRn0.ps} }

\caption{{\it Upper panel}: the probability distribution $P(N)$ for
counts in cells as in Fig.\ref{figPN} for $n=0$ with the constraint
$800 < \xia <1600$. We show the curves obtained for the three comoving
scales $R=0.5$ Mpc (triangles), $R=1$ Mpc (squares) and $R=2$ Mpc
(circles). The solid curve is the analytical fit given by (\ref{fithx})
and Tab.2 as in Fig.\ref{figPN}. {\it Lower panel}: the curves for
$P(N)$ obtained at fixed comoving $R$ by letting $\xia$ (i.e. time)
vary. We show the cases $R=1$ Mpc (triangles) and $R=4$ Mpc (squares).
Different shades of grey correspond to different values of $\xia$.}
\label{figPNtest}

\end{figure}

Although all curves at different scales superpose, building indeed the
predicted universal $h(x)$ curve, some deviations are seen for small
values of the parameter $x$. Clearly, these small values of $x$
($\log(x) = -2.5$ !), never reached in earlier simulations, are at the
limit of the possibilities of the present calculations, but the
deviations appear to be systematic. For larger $\xia$ the function
$h(x)$ appears to be slightly flatter although it still obeys the
constraints (\ref{hnorm}). Whether this effect is the sign of a real
deviation of $h(x)$ from the predicted scaling, or is due to some bias
in the simulation, warrants investigation. To this purpose, we can
consider the same simulation results in more detail. In the continuous
limit ($N \gg 1$), the expression $N^2/\Na \; P(N)$ that in general
depends on the three variables $N$, the cell size $R$ and the expansion
parameter $a$, can always be written
as $N^2/\Na \; P(N) = x^2 h(x, \xia,R)$, the latter expression
representing a simple change of the original variables into $x$, $\xia$
and $R$. The exact
self-similarity due to our power-law initial conditions implies that
$N^2/\Na \; P(N)$, as well as $\xia$, are a function of the ratio
$R/ a^{(n+5)/(n+3)}$ only (i.e. they only depend on $\sigma$) and thus
that
$N^2/\Na \; P(N) = x^2 h(x,\xia)$ is independent of $R$ since $\xia$
carries all the dependence on $R$ and $a$.
The scaling prediction (due to the stable-clustering model) in addition
states that in the limit of large $\xia$, when the stable clustering
regime is reached, $N^2/\Na \; P(N)$ has a limit for fixed $x$ that is
independent of $\xia$. By definition of $h(x)$ this limit is $N^2/\Na
\; P(N) = x^2 h(x)$. The scatter seen in Fig.\ref{figPN} must have its
origin in the violation of one of the two above conditions that are
needed to get the seeked scaling function. As a first check, we can
consider the variations with $R$ at fixed $\xia$. The value of the
latter ranging from 100 to 6400, typical intermediate values are $800
\le \xia \le 1600$. In this range, values of $R=0.5, 1$ and $ 2$ Mpc
are available. We can see in Fig.\ref{figPNtest} (upper panel) that for
these given values of $\xia$, although the curves look reasonably
close, there is a distinct drift at small $x$, the curves for the
smaller values of $R$ lying systematically above the ones for the lower
values. Also, the 
 deviations get larger the smaller the value of $x$. Since we let $R$
vary for a fixed $\xia$ this effect represents deviations from the
``exact'' self-similarity (implied by scale-invariant initial
conditions) due to the unavoidable errors in the simulation.
To estimate the size of these deviations, we show in
Fig.\ref{figPNtest} (lower panel) the probability distribution obtained
for all possible values of $\xia$ keeping $R$ fixed at two values (1
and 4 Mpc). For a given $R$, when $\xia$ is varied all curves
superpose. But when $R$ is changed, we get a different curve. This
shows again that $N^2/\Na \; P(N)$ depends on $R$ and not on $\xia$,
with a deviation that reaches a factor of 2 at small $x$ ($x < 3 \;
10^{-3}$), of the order of magnitude of the scatter seen in
Fig.\ref{figPN}. So we can attribute the observed scatter to deviations from the ``exact''self-similarity (due to a power-law initial power-spectrum), hence to
numerical inaccuracies and not to a violation of the non-linear 
scaling prediction that we aim to test here.

\begin{figure}

\centering \psfig{figure=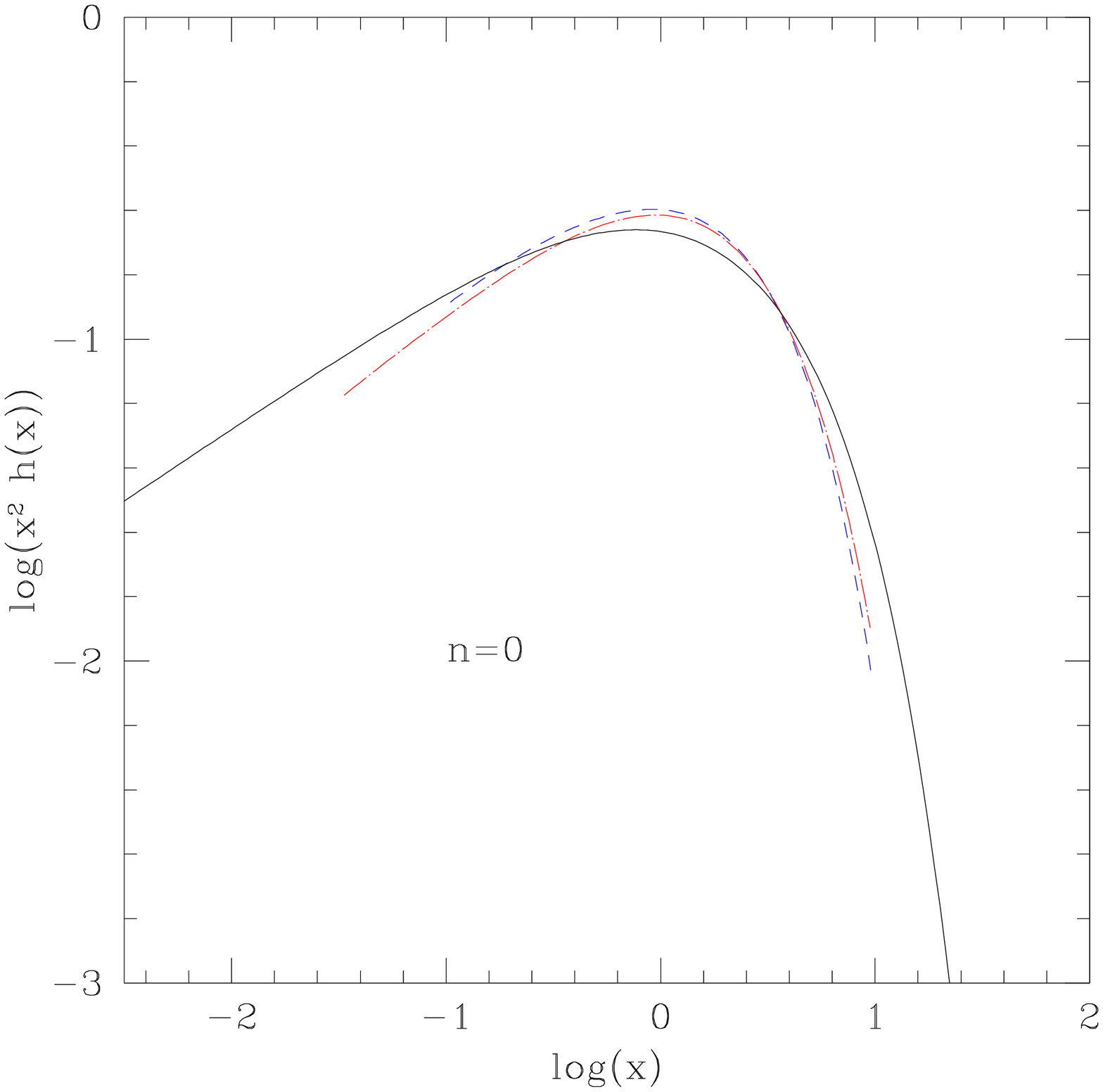,width=8cm,height=6cm}
\centering \psfig{figure=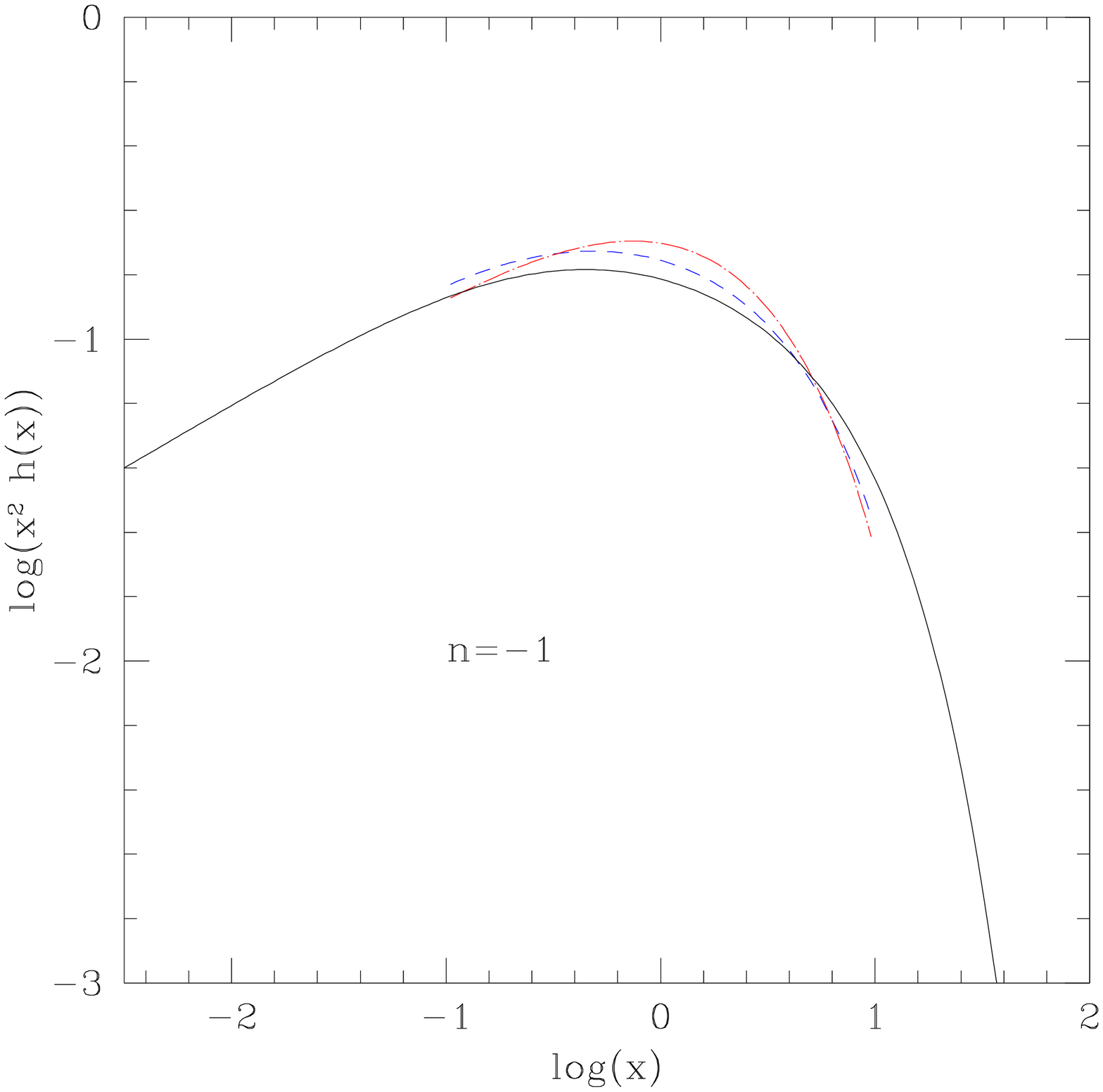,width=8cm,height=6cm}
\centering \psfig{figure=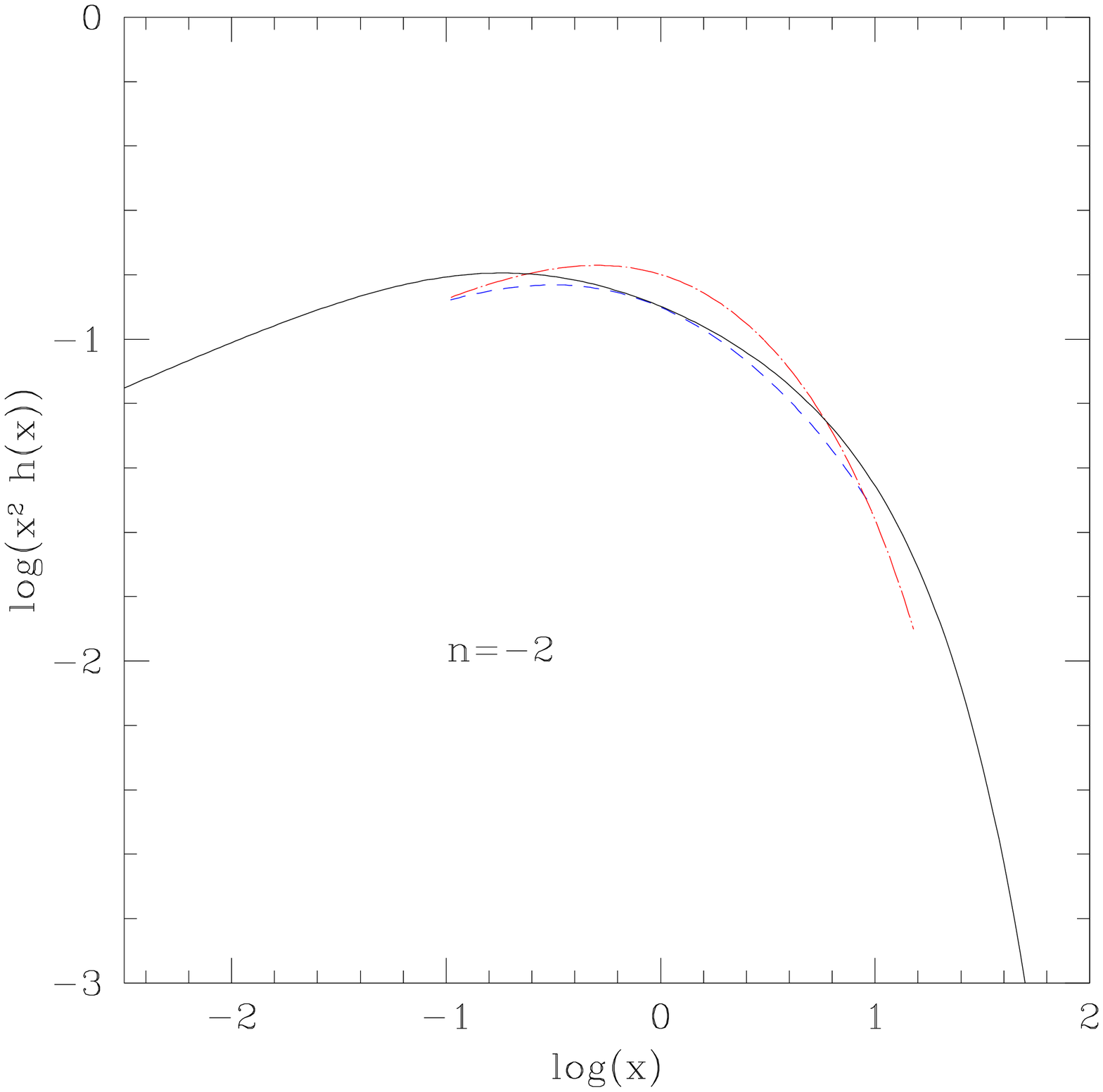,width=8cm,height=6cm}

%{\epsfxsize=8 cm \epsfysize=5.5 cm \epsfbox{fighxn0.ps} }
%{\epsfxsize=8 cm \epsfysize=5.5 cm \epsfbox{fighxnm1.ps} }
%{\epsfxsize=8 cm \epsfysize=5.5 cm \epsfbox{fighxnm2.ps} }

\caption{Comparison of the scaling function $x^2 \; h(x)$ obtained from
our
simulations with the estimates given by Colombi et al.(1997) (dashed
line) and Munshi et al.(1999) (dot-dashed line). For the latter two
cases we only display $x^2 \; h(x)$ in the range where it was actually
tested against numerical data.}
\label{fighxsim}

\end{figure}

We compare in Fig.\ref{fighxsim} the scaling functions $h(x)$ we get
in our simulations with the results obtained by Colombi et al.(1997)
and Munshi et al.(1999). For these two latter cases we only display
$h(x)$ over the ranges where there were actually data points from the
simulations (for our scaling function the range tested against
numerical results can be directly seen from Fig.\ref{figPN}).
We can check in the figure that all of the scaling functions agree
fairly well with each other (within the numerical uncertainties which
can be
estimated from the dispersion shown in Fig.\ref{figPN}). Thus,
although there is still some dispersion between the various estimates
of $h(x)$, which could be expected from the discrepancies which
already appeared for $\xia$, the scaling function $h(x)$ is rather
well constrained in the range where it has actually been tested, to an
accuracy which is certainly sufficient for practical
purposes. However, it should be noted that the range in $x$ available
to constrain $h(x)$ is still too small to determine accurately the
parameters of the fit (\ref{fithx}) for low $n$, especially $x_s$
(there is some degeneracy between the exponential cutoff and the
exponent $\omega_s$ of the power-law pre-factor). Thus, in the case
$n=-2$ the values of $x_s$ obtained by Colombi et al.(1997) and Munshi
et al.(1999) differ by a factor 2, although both scaling functions are
rather close over the range shown in Fig.\ref{fighxsim}. This means
that
the asymptotic behaviour of $h(x)$, and more generally its behaviour
for $x \ge 10$, typically, is still rather poorly determined.

\subsection{Moments of the counts in cells }
\label{Moments of the counts in cells}

If one believes that the scaling function $h(x)$ provides a reasonably
good description of the counts-in-cells, it is readily seen from
Fig.\ref{figPN} that the measured counts in the simulation miss the
very rare overdensities at $x \ge 10$. In this region, actually there
are counts that oscillate, due to (Bouchet et al.1992; Colombi et
al.1995) whatever last cluster appears to be in the numerical data,
reflecting their ``cosmic variance''. It turns out that the calculation
of the moments $S_p$ (\ref{Sp}) badly reflects this missing
information. The moments directly obtained from the measured counts
are systematically underestimated, the effect being more pronounced
as the scale of the cells increases. The moments depend most on the
counts-in-cells at $x \simeq (p + \omega_s) x_s$, as can be seen from
(\ref{Sphx}) and (\ref{has}). With $\omega_s \simeq -1$ and $x_s \simeq
10$ the calculation of $S_3$, already, cannot be done without an
extrapolation of the numerical data 
 to the badly needed, but not
 measured large $x$ (large $N$) tail. Appropriate methods
have been proposed by the above authors
, but the resulting values of
$S_p$ extracted from the data strongly reflect the precise
extrapolation procedure used by the various authors (Colombi et
al.1996; Munshi et al.1999)
\footnote{
To increase the available range, Colombi et al (1996)
 correct for the oscillations (typically a factor of 3 from the mean, sometimes an order of magnitude) due to
this ``cosmic variance''
by means of an  an educated guess, fitting a smooth curve of pre-determined shape
to the data.
 To estimate the uncertainty,
 these authors determine by human judgement what can
be considered as a reasonable fit and what cannot. This generates 
error bars given in the form of a factor $f_Q^{\pm1}$ (i.e. $S_p$ is
within the range $S_{p,mean}/f_Q$ to $S_{p,mean} \times f_Q$) with
typically
$f_Q = 1.1, 1.35, 1.65$ for $p = 3,4,5$, respectively, whereas the
deviations from scaling show, under the same conditions, 
a scale-dependence by a
factor $f = 1.2, 1.5, 1.85$, respectively. Their interpretation 
is that, since  $f$ is typically  larger than $ f_q$ by $10 \%$, this
`` prooves  a small but significant departure from the stable clustering
predictions'' in the form of a scale-dependence of the coefficients $S_p$. 
Whereas we think that such a procedure (see Colombi et
al.1994)
to extract information on the large $N$ behaviour, although somewhat
indirect, is reasonable in the absence of any better way, 
it is not free from systematic biases and the
associated correction may not be achieved at the accuracy 
needed to back the claim of
the former authors.
Indeed, a subsequent work (Munshi et al.1999), using a different
computation, shows that the numerical data for large $x$ are
at least  consistent with the assumption of
no scale-dependence at all of the coefficients $S_p$.
}
.
The differences get rapidly very large
with increasing order $p$, from factors of 2 for $p = 4$ to already an
order of magnitude for $p=5$. Thus extreme care must be taken in the
interpretation of the $S_p$ parameters extracted from the simulations.
Rather than introducing sophisticated corrections to extract the large
$N$ tail of the counts (with uncontrolled errors due to the badly
needed correction for cosmic variance) it may be as accurate
(inaccurate) to simply fit a function $h(x)$ to the numerical data and
use (\ref{Sphx}) to get the moments.

\section{Mass functions}
\label{Mass functions}

For astrophysical applications, one is usually more interested in the
mass functions (the number density of objects defined for instance by
a given density threshold) than in the counts in cells
statistics. Moreover, one may wish to study a wide variety of
objects, from low-density Lyman-$\alpha$ clouds (which may even be
underdense!) to clusters and massive very dense galaxies (see for
instance Valageas \& Schaeffer 1999 and Valageas et al.1999 for a
detailed description of Lyman-$\alpha$ clouds and galaxies in this
framework). Of course, these very different objects are defined by
specific constraints  so that they cannot be described solely by
the mass function of ``just-collapsed'' halos with the traditional
overdensity $\Delta = 177$.

We have determined numerically the multiplicity of various kinds of 
idealized objects and compared them to theoretical predictions.
To this purpose, we use two different methods. The fist is  the  ``spherical
overdensity'' algorithm (Lacey \& Cole 1994) which ranks particles in
order of decreasing density and counts halos defined by the density
contrast $\Delta$ by looking down this list (this introduces some double counting since part of a halo may be counted again at a lower rank:
to avoid this,  when a new halo
is recognized its particles are removed from the list). The second
method is based on  a friend-of-friend algorithm with a
linking length $b$ which we relate to $\Delta$ through:
\beq
(1+\Delta) = 178 \; \left( \frac{0.2}{b} \right)^{3} .
\eeq

\subsection{The Press-Schechter approximation for just-collapsed halos}
\label{The Press-Schechter approximation for just-collapsed halos}

\begin{figure}

\centering \psfig{figure=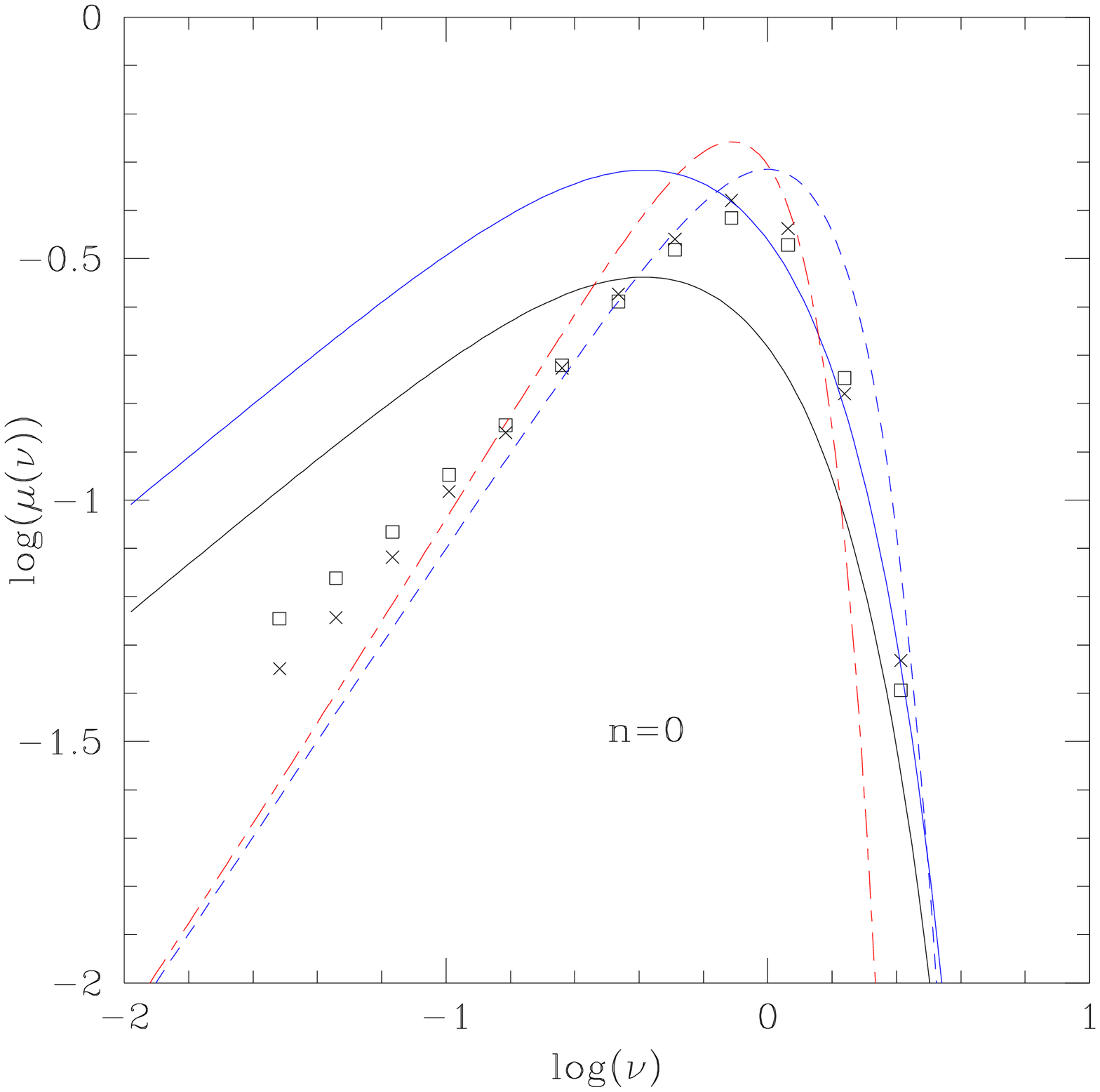,width=8cm,height=6cm}
\centering \psfig{figure=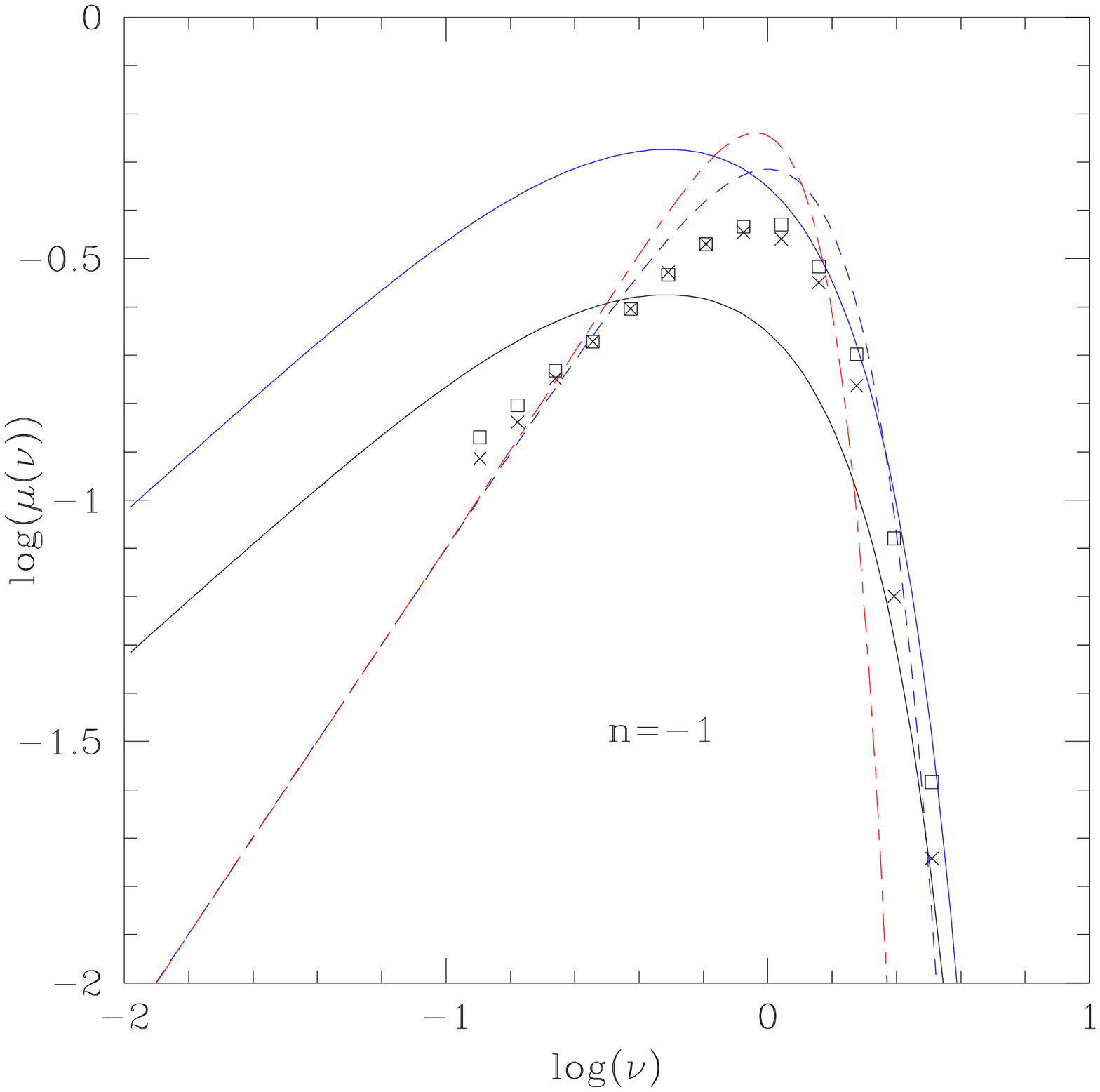,width=8cm,height=6cm}
\centering \psfig{figure=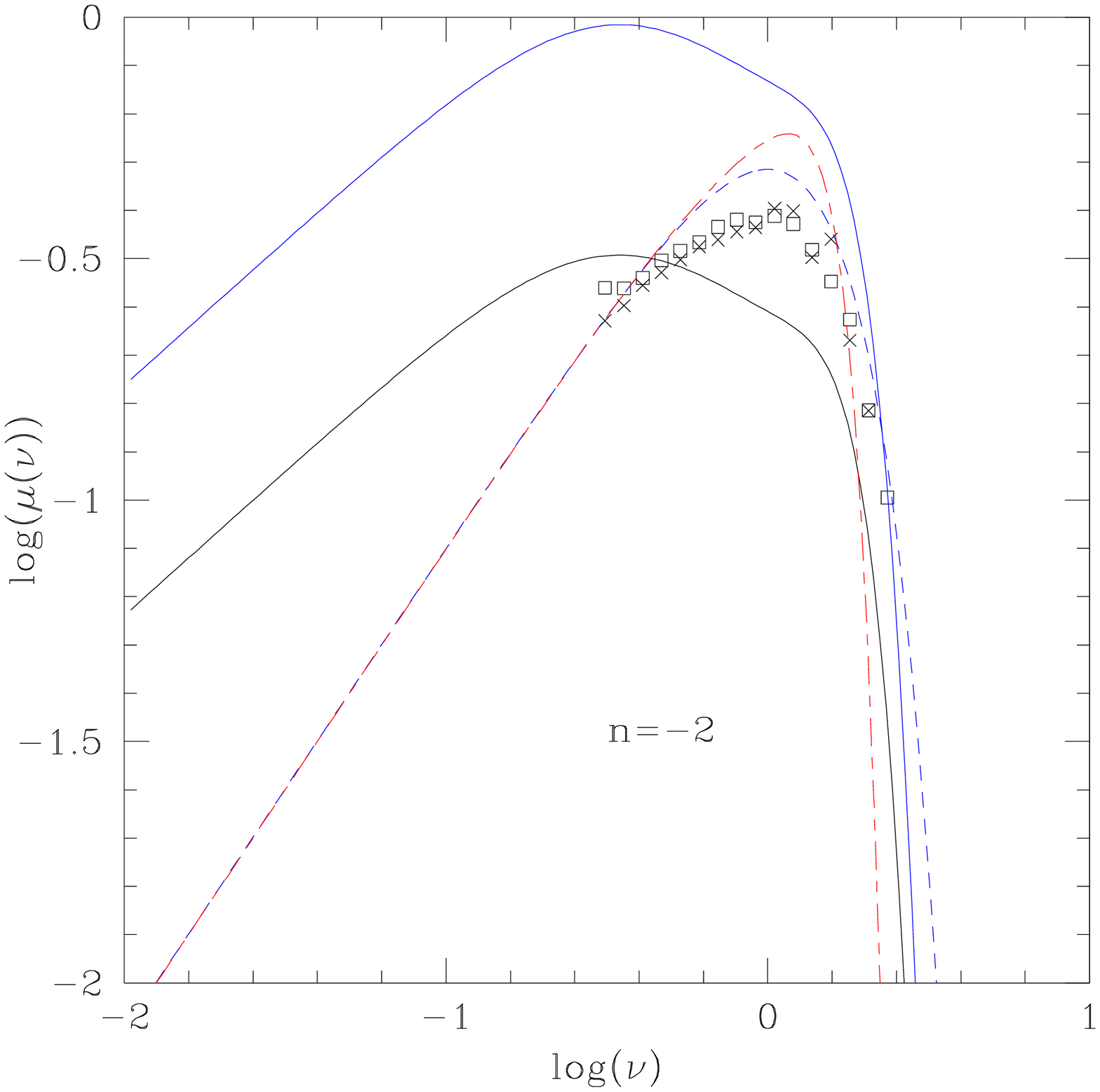,width=8cm,height=6cm}

%{\epsfxsize=8 cm \epsfysize=5.4 cm \epsfbox{figPSn0.ps} }
%{\epsfxsize=8 cm \epsfysize=5.4 cm \epsfbox{figPSnm1.ps} }
%{\epsfxsize=8 cm \epsfysize=5.4 cm \epsfbox{figPSnm2.ps} }

\caption{The mass function of ``just-collapsed'' halos defined by the
density threshold $\Delta=177$, displayed in terms of the variable
$\nu$. 
The crosses show the results of a ``spherical overdensity''
algorithm while the squares correspond to a ``friends-of-friends''
procedure. The numerical results are averages over
different output times from the simulation. 
The short-dashed curve is
the standard PS prescription (note it does not depend on $n$). The dot-dashed
line corresponds to (\ref{hPSs}),  referred to as PSs in the text,
which is the same as the PS result in the case that the
stable clustering regime
 is reached (and it can be seen that this is
not
fully the case here). The lower solid line is the scaling
function $h(x)$ obtained from counts-in-cells while the upper solid
line is $3/\gam \; h(x)$. The scaling model predicts that in
the stable clustering regime the counts are in-between these two curves
at large $\nu$, and below the upper one but with generically the same
slope at small $\nu$.}

\label{figPS}

\end{figure}

Most studies have focused on the mass function of ``just-collapsed''
(or ``just-virialized'') halos which are defined by a density contrast
in the actual non-linear density field $\Delta \simeq 177$ (for
$\Omega=1$). Indeed, this is the quantity which is considered by the
popular Press-Schechter prescription (Press \& Schechter 1974). The
idea of this formulation is to recognize in the early universe, when
density fluctuations are still gaussian and described by the linear
theory, which overdensities will eventually collapse and form
virialized objects. Thus, one associates to any linear density
contrast $\delta_L$ obtained by the linear theory ($\delta_L \propto
a$ for a given fluctuation of fixed mass) a non-linear density
contrast $\Delta$ provided by the spherical model. Next one assumes in
addition that collapsed objects virialize at a radius equal for
instance to one half their turn-around radius at the time when the
spherical dynamics predicts a singularity. Then one obtains that when
$\delta_L$ reaches the threshold $\delta_c \simeq 1.69$ a virialized
halo with a density contrast $\Delta \simeq 177$ has formed. Finally,
one identifies the mass fraction $F_{\Delta}(>M)$ within such objects
of mass larger than $M$ with the fraction of matter which is above the
threshold $\delta_c$ at scale $M$ in the linear universe:
\beq
F_{\Delta}(>M) = F_L(>\delta_c;M) = \int_{\delta_c}^{\infty}
 \frac{d\delta}{\sqrt{2\pi} \sigma(M)} \; e^{-\delta^2/(2\sigma^2)}
\eeq
Then, the mass-derivative of the previous quantity provides the mass
function of virialized halos:
\beq
\mu(M) \frac{dM}{M} = - 2 \; \frac{dF_{\Delta}}{dM} \; dM
= \sqrt{\frac{2}{\pi}} \; \nu \; e^{-\nu^2/2} \;
\frac{d\nu}{\nu}  \label{PSmf}
\eeq
with
\beq
\nu = \frac{\delta_c}{\sigma(M)}
\eeq
Here we defined $\mu(M) dM/M$ as the fraction of matter which is
enclosed in objects of mass $M$ to $M+dM$. Note that in (\ref{PSmf})
we  have arbitrarily
multiplied the mass function by the usual factor 2 so as to get the
proper normalization to unity. We must however emphasize that this
normalization correction is not justified in the present case. Indeed,
although this multiplicative factor 2 was recovered rigorously by Bond
et al.(1991) using the excursion sets formalism for a top-hat in $k$,
taking into account the cloud-in-cloud problem, this result does not
apply to more realistic filters like the top-hat in $R$ used here. In
fact, as shown in VS (and noticed by Peacock \& Heavens 1990 through
numerical results) this correction factor goes to unity at large
masses. So, this factor is not constant (and equal to 2)
but $\nu$-dependent. This simple normalization procedure
(\ref{PSmf}), nevertheless, leads to a scaling-law in the parameter
$\nu$: the mass function $\mu(\nu) d\nu/\nu$ does not depend any longer
on the initial power-spectrum. We present in Fig.\ref{figPS} the
comparison of our numerical results with the PS
prescription. 
We can check in the figure that the mass functions obtained by both
methods are consistent. Note that the numerical points correspond to
averages over different output times (weighted by the number of
halos) of the mass functions realized in the simulations. Although
there is some scatter we checked that all curves superpose on the mean
mass function shown in Fig.\ref{figPS}. We can note that our results
are consistent with the scaling in $\nu$ obtained from the PS
approach.

Thus, we can see as was already checked by many authors
(e.g. Efstathiou et al.1988; Kauffmann \& White 1993; Lacey \& Cole
1994) that the Press-Schechter mass function (shown by the dashed
line) agrees reasonably well with the numerical results. Indeed,
although there are some small but significant discrepancies
(especially at the low mass end where the slope 
seems to tend towards the non-linear prediction).
This could be expected in view of
the simplicity of this approach but the agreement is still rather good for
such a simple model. A drawback of the Press-Schechter  approach is
that it is clearly limited to the mass function of ``just-collapsed''
objects and  does not provide by itself predictions for mass
functions defined at other density contrasts.

\subsection{The scaling model for the halo multiplicity}
\label{The scaling model for the halo multiplicity}

Predictions for the mass function of objects defined by an arbitrary
density contrast $\Delta$ in the actual non-linear density field can
be obtained within the framework of the scaling model, as shown in
detail by VS. These predictions are valid for {\it any} value of
$(1+\Delta)$, whatever large or small (even below unity!), provided the
conditions (\ref {valhx}) are met. As was the case for the counts in
cells the mass function exhibits a scaling-law in the parameter $x$:
\beq
\mu(M) \frac{dM}{M} = x^2 \; H(x) \; \frac{dx}{x}  \label{muHx}
\eeq
where we still have $x=(1+\Delta)/\xia(R)$. Note that here $R$ is the
actual size of the halo (as opposed to the ``mass scale'' $R_L$, the
radius an object of the same mass would have in a uniform density
universe, which enters for instance in the PS mass function). The
scaling function $H(x)$ should be related to $h(x)$ introduced for the
counts in cells. Indeed, if one identifies the mass within objects
(defined by the density threshold $\Delta$) larger than $R$ (i.e. more
massive than $M=(1+\Delta) \rhoa (4 \pi/3) R^3$) with the fraction of
matter enclosed in cells of scale $R$ with a density contrast larger
than $\Delta$, one obtains a simple estimate $H_{cell}$ of $H(x)$
which reads
\beq
H_{cell}(x)=h(x).
\label {Hcell}
\eeq
Even if this is probably an oversimplified approach, 
it has been steadily argued (Balian and Schaeffer 1989a ; Bernardeau and Schaeffer 1991; VS)
at an increasing level of sophistication 
that the functions $H(x)$ and $h(x)$ can be expected
to be  close and that for practical purposes $h(x)$ can be used as a good approximation to $H(x)$. 
Also, as can be seen in VS, this procedure is found
to be very close to the global derivative used in the PS prescription (\ref{PSmf}).
In the scaling model, however, it is applied to the actual non-linear
density field rather than to the original Gaussian field,
and thus is correctly normalised.

 On the other
hand, if one defines directly objects of scale $R$ to $R+dR$ by the
constraints that the density contrast is larger than $\Delta$ on scale
$R$ but smaller than $\Delta$ on scale $R+dR$ (thus one tries to
recognize {\it individually} each halo in order to handle the
cloud-in-cloud problem) one gets a new scaling function $H_{><}(x)$
which can be shown (VS) to satisfy (for constant $\Delta$):
\beq
\left\{  \begin{array}{rl}    \forall x : & {\displaystyle H_{><}(x)
\leq
\frac{3}{\gam} \; h(x) }  \\  \\   x \gg x_s : & {\displaystyle
h(x) \leq H_{><}(x) \leq \frac{3}{\gam} \; h(x) } \end{array} \right.
\label{H><}
\eeq
with generically (unless the leading order in the expansion
accidentally vanishes) the same asymptotic behaviour at small $x$:
\beq
x \ll 1 \; : \hspace{1cm} H_{><}(x) \propto h(x)  \propto   x^{\omega -
2}
\label {Hlowx}
\eeq
Moreover, for a tree-model where the many-body correlation functions
can be expressed as products of the two-point correlation function one
obtains $H_{><}(x) = h(x)$ for $x \gg x_s$. Note that for a gaussian random field the same procedure (that is to define the mass function by requiring a given overdensity in the linear regime at a given scale $R$ and a lower one at a slightly larger scale, in order to recognize individually each object), would lead to a mass function which diverges at small masses (see VS), meaning divergent total mass due to the severe overcounting.
Even for a fixed  density contrast $\Delta$,  the same mass is  counted many times at smaller and smaller scales because of numerous density inversions. Since the fraction of matter (compared to unity) enclosed in objects defined by a density threshold is a direct measure of the severity of the ``cloud-in-cloud'' problem, this shows the latter is indeed severe for Gaussian fluctuations. On the other hand, in the non-linear case when one directly counts the overdensities in the non-linear density field, the normalization is necessarily lower than $3/\gam$ for the case described above from (\ref{H><}) and can be expected to be close to unity.
This shows, as argued by VS, that in the non-linear regime, the
overdensities being well-defined, there is no longer any serious
cloud-in-cloud problem. The latter still manifests itself in the sense
that the use of $H_{><}$ induces some slight overcounting,
typically of the order of $20\%$ (see VS). This is the same overcounting as the one found in computations for the spherical density algorithm
 (which has been  removed in this case by attributing all the mass to the larger object), and indeed of similar magnitude,
 as we checked in the
numerical simulations. The true constraint, rather than
the one on the edge of the halos used to define $H_{><}$, which
avoids this overcounting, is the condition that the density contrast
is smaller than $\Delta$ for {\it all} scales larger than $R+dR$.
This would completely solve the cloud-in-cloud problem. However, the
previous formulation leading to $H_{><}(x)$ should already provide a
satisfactory tool. Moreover, it is close to the algorithm actually
used in numerical simulations.

The solid curves in Fig.\ref{figPS} represent the functions $h(x)$ and
$3/\gam \; h(x)$. From the results described above, we expect the mass
function obtained from the numerical simulations to lie below the
upper curve for all $\nu$ and between these two curves for large $\nu$
(which corresponds to large $x$) beyond the exponential cutoff. We can
see that this is indeed approximately the case for all power-spectra.
However, it appears that the numerical mass function is significantly
different from $h(x)$, especially at the low mass end. This could be
expected since it does not seem possible to derive a lower bound for
$H(x)$ in this domain in the general case. Nevertheless, the slopes of
both scaling functions appear to be the same $H(x) \propto h(x)$ for
$x \ll x_s$. Note that for large $\nu$ and $x$ ($x > \Delta/10$) the
mass function should not behave as (\ref{muHx}) since this domain
corresponds to large scales which are no longer in the non-linear
regime, so the correlations functions do not obey the scaling
(\ref{scal1}). The dot-dashed lines correspond to the ``PSs'' mass function
(\ref{muHx}) given by the scaling approach with $H(x)$ taken equal to
the ``spherical model'' scaling function (\ref{hPSs}). Of course at
small scales it superposes onto the standard PS mass function, as explained in Sect.\ref{Counts in cells statistics}, since in this regime it is equivalent to the PS prediction. At large scales
it differs from the latter because one leaves the highly non-linear,
stable-clustering,
regime and the two-point correlation function is no longer given by
the asymptotic behaviour $\xia(R) = [10/(3\alpha) \; \sigma(R_L)]^3$,
see (\ref{Fasym}). Note that the PS approach works somewhat better for
the
mass function than it did for the counts in cells statistics, and that,
for the contrast  $\Delta = 177$ relevant to the present calculations,
at large masses the difference with respect to the stable clustering PSs result somewhat helps.
What helps also to bring the PS result in agreement with the numerical data is the factor of $2$ by which this mass function has been multiplied, as is traditional,
despite numerical calculations (Peacock \& Heavens 1990) as well as 
theoretical considerations (VS) show this factor should be close to unity 
at large $x$ (and is expected to be larger at small $x$).

As compared to the standard  PS approach, the interest of the formulation
(\ref{muHx}) for the mass function of ``just-virialized'' objects is
that it makes a connection with another characteristics of the density
field: the counts in cells statistics described earlier.

\subsection{Different constant density contrasts}
\label{Different constant density contrasts}

The main advantage advantage of the scaling model is that it can deal with more
general mass functions. 
We shall first consider the case of mass functions which are still
defined by a constant density contrast but where $\Delta$ can now take
any
value. In the scaling approach which led to (\ref{muHx}) the quantity
$(1+\Delta)$ is only used to define objects but no specific value is
singled out as opposed to the PS mass function which explicitly deals
with ``just-collapsed'' objects (so that it makes sense for only one
class of halos characterized by a non-linear density contrast $\Delta
\sim 177$). Hence the relation (\ref{muHx}) should hold for any value
of $(1+\Delta)$, in the domain where the scaling (\ref{PNhx})
holds.

\subsubsection{Spherical overdensity agorithm}
\label{Spherical overdensity agorithm}

\begin{figure}

\centering \psfig{figure=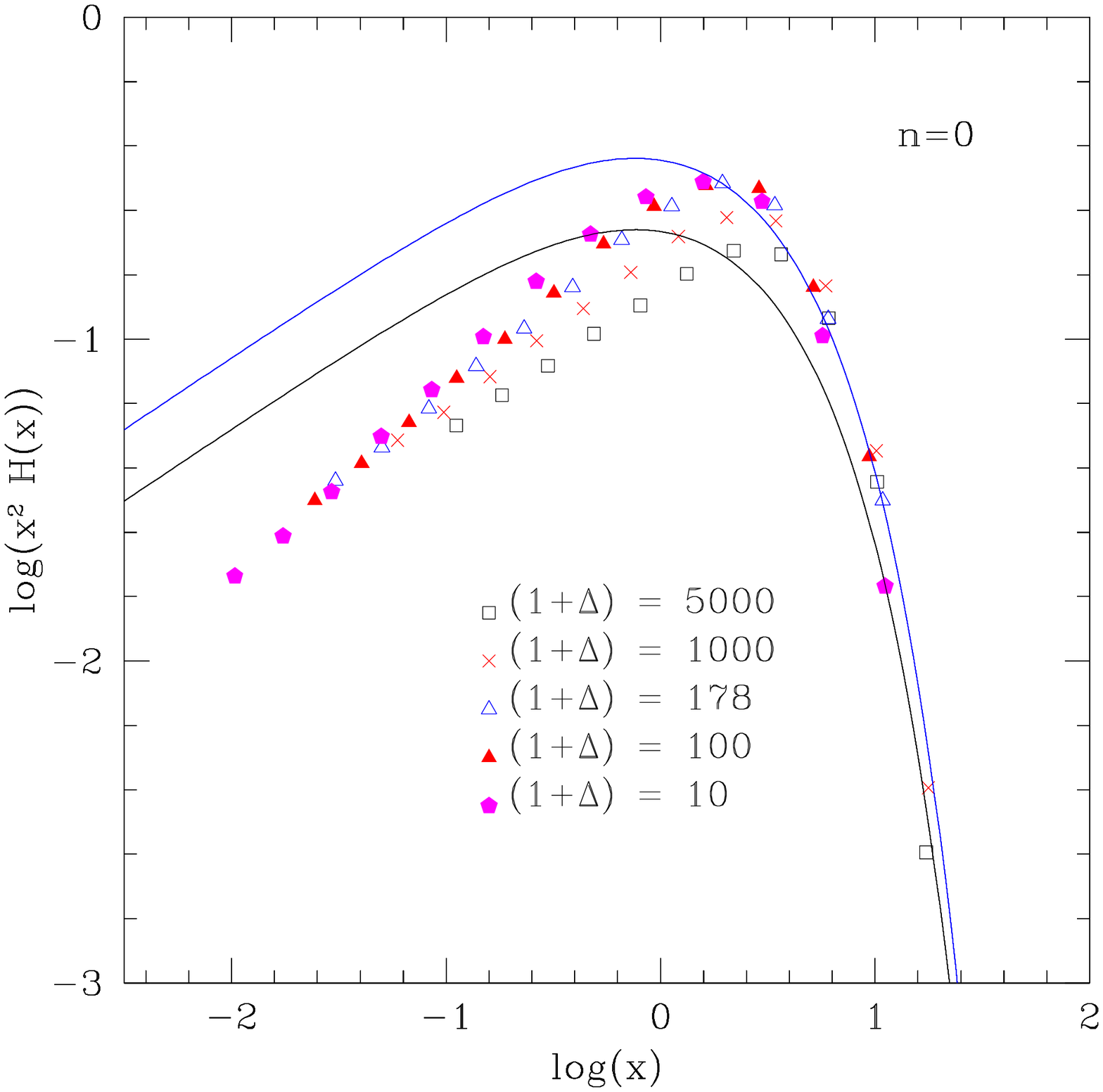,width=8cm,height=6cm}
\centering \psfig{figure=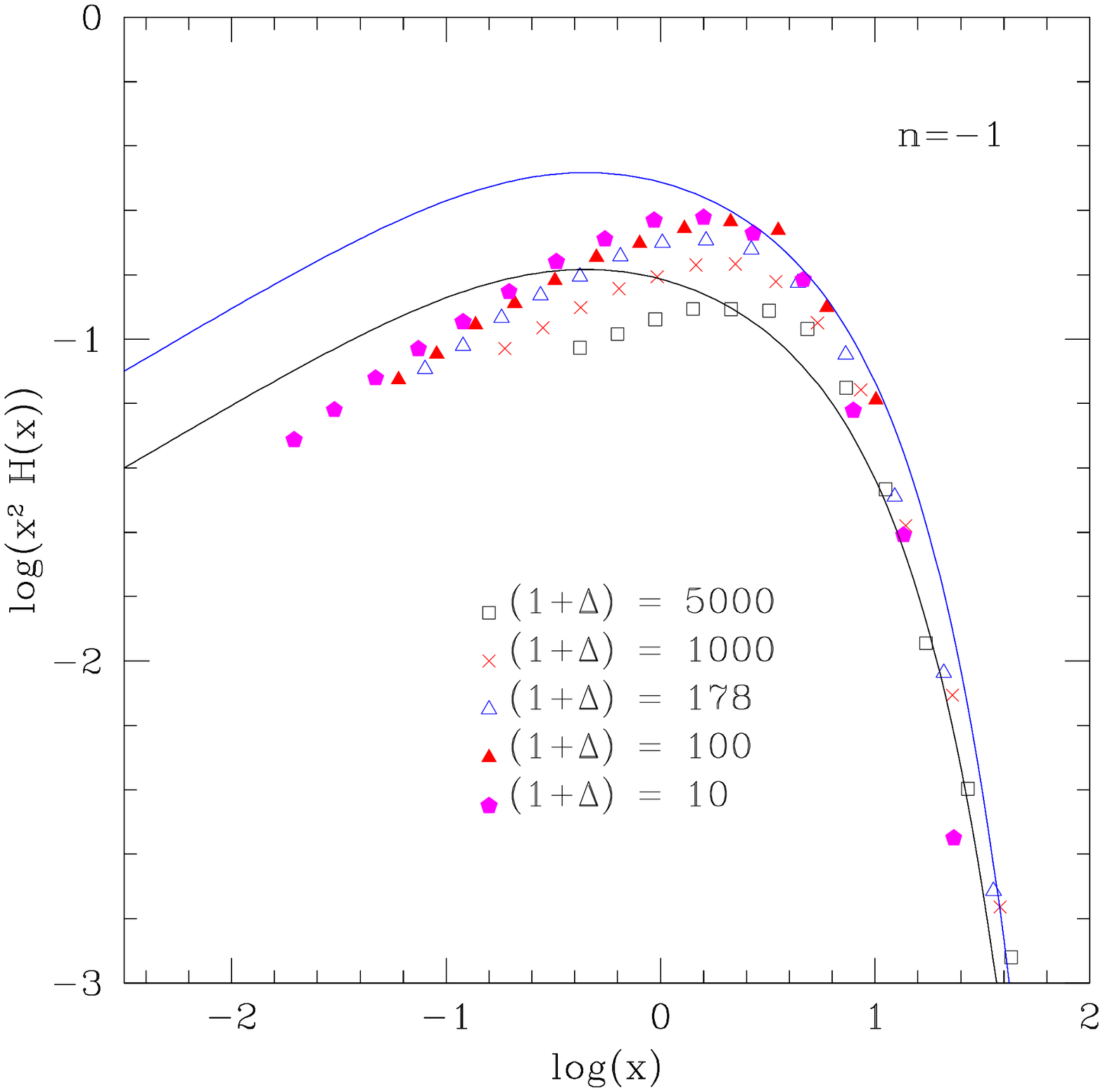,width=8cm,height=6cm}
\centering \psfig{figure=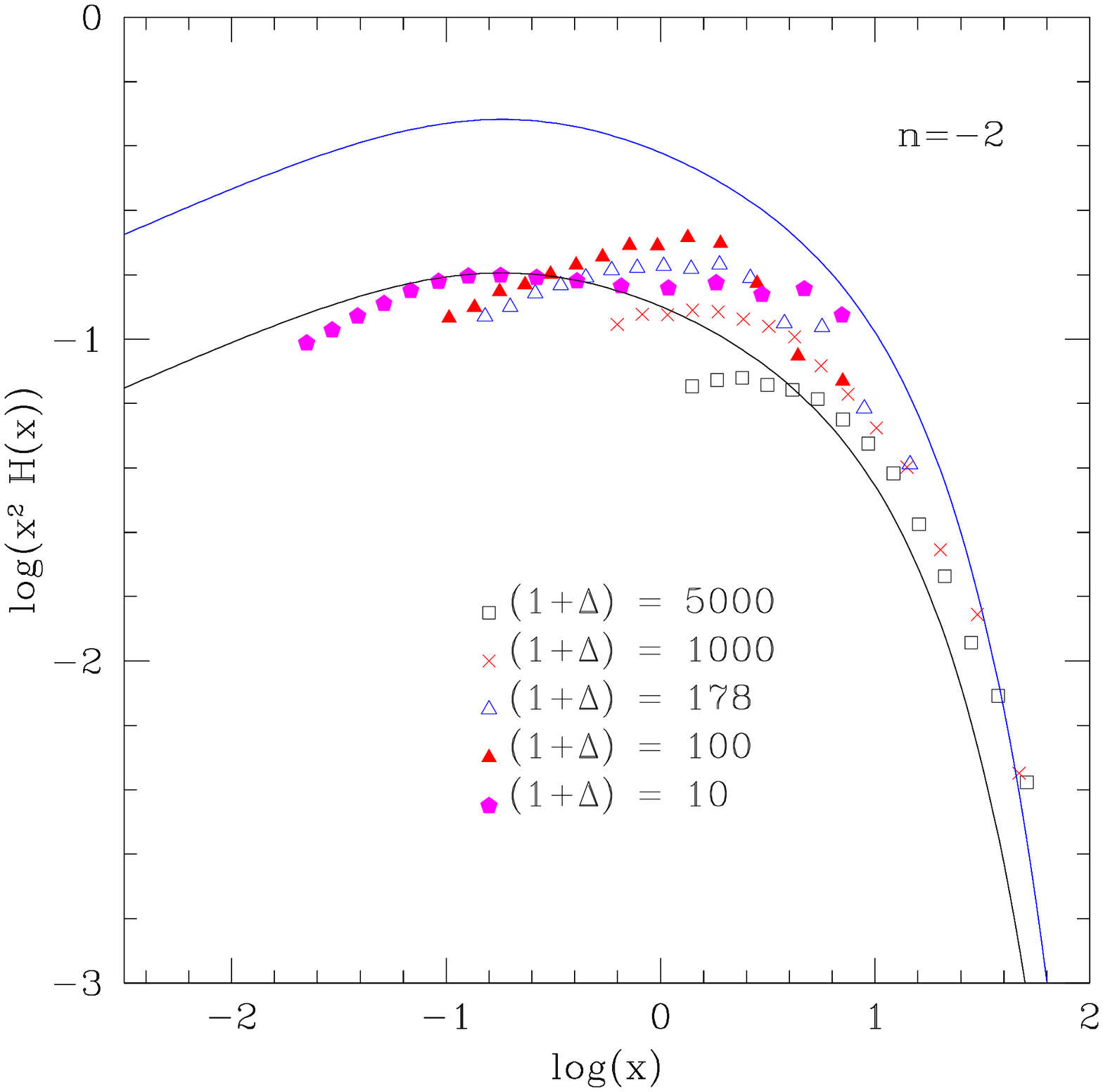,width=8cm,height=6cm}

%{\epsfxsize=8 cm \epsfysize=5.5 cm \epsfbox{figDelson0.ps} }
%{\epsfxsize=8 cm \epsfysize=5.5 cm \epsfbox{figDelsonm1.ps} }
%{\epsfxsize=8 cm \epsfysize=5.5 cm \epsfbox{figDelsonm2.ps} }

\caption{The mass functions of halos defined by various density
thresholds
$\Delta$ obtained from a ``spherical overdensity''
algorithm. Different symbols correspond to different values of
$(1+\Delta)$. The lower solid curve is the scaling function $h(x)$
measured from counts in cells while the upper solid curve is $3/\gam
\; h(x)$. The non-linear model predicts that
the counts are in-between these two curves at large $x$, and
below the upper one but generically with the same slope for small $x$.}

\label{figDelso}

\end{figure}

We show in Fig.\ref{figDelso} the mass functions we
obtain from the numerical simulation for various values of the density
contrast for the ``spherical overdensity'' algorithm. We display the
quantity $\mu(M) \; dlnM/dlnx$ as a function of $x=(1+\Delta)/\xia(R)$
(as explained previously $\xia(R)$ was obtained from the same
simulation through counts in cells, see Fig.\ref{figXisigR}). If the
scaling (\ref{muHx}) holds
all curves should superpose (in the relevant range of validity of the
scaling laws) onto the function $x^2 \; H(x)$ (we only show the range
$x < (1+\Delta)$ which corresponds to $\xia > 1$ but the range of
validity could be smaller than this: $\xia \gg 1$). The points
correspond again to an average over different output times.

We can see that the available counts are pushed much further towards
the larger values of $x$ as compared to Fig.\ref{figPN}. This is
because we check the neighbourhood of particles to get the halos and
thus we are guaranteed to be in a region where there really is some
mass. To get the statistics of  the counts-in-cells, on the other hand,
implies to check randomly over the whole volume, so that the statistics
of the less dense but more extended regions are relatively favoured. As
will be discussed in more detail in Sect.\ref{Constant radius}, this
amounts in some sense to measure $xH(x)$ rather than $H(x)$, with a
better statistics at the large values of $x$ (note, however, that
whereas $x^2H(x)$ has a maximum, both the above functions are
decreasing for increasing $x$, still favoring the small values of $x$
as compared to the large ones). A similar effect was described in
Sect.\ref{The two-point correlation function} for the measure of $\xit$
as opposed to $\xia$.

As expected, all curves lie between $h(x)$ and $3/\gam \; h(x)$ beyond
the exponential cutoff and are located below $3/\gam \; h(x)$ for all
values of $x$. Moreover, the slope at small $x$ is in most cases close
to that for
$h(x)$, although for $n=0$, where the statistics are the best, a distinct difference is apparent: whereas  $h(x)$ has a slope (Tab. 2) $\omega = 0.45$, here the slope is rather $\omega = 0.6$.

Moreover, although all curves superpose beyond the cutoff as
predicted, there appears to be a shift in the power-law domain with
$(1+\Delta)$: larger density contrasts lead to a
smaller amplitude in this part of the mass function. This drift appears
at $x \leq 3$ for $(1+\Delta) = 5000 $ and $x \le 1$ for $(1+\Delta) =
1000$ and has disappeared for $(1+\Delta) = 178$ (indeed the curves
obtained for $10 \leq (1+\Delta) \leq 178$ superpose very well).

These deviations will be discussed in the following section.

\subsubsection{Friend-of-friend algorithm}
\label{Friend-of-friend algorithm}

\begin{figure}

\centering \psfig{figure=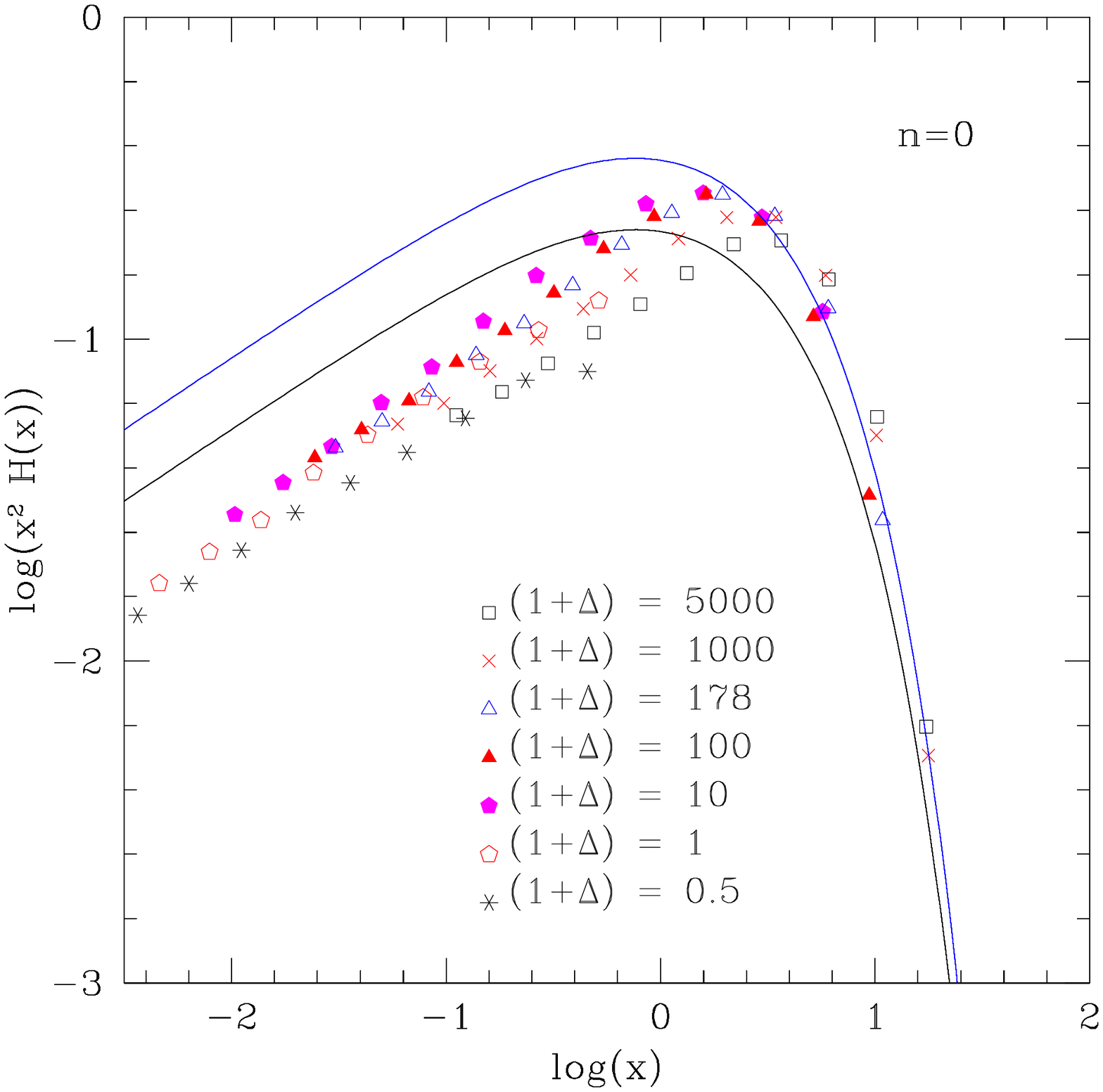,width=8cm,height=6cm}
\centering \psfig{figure=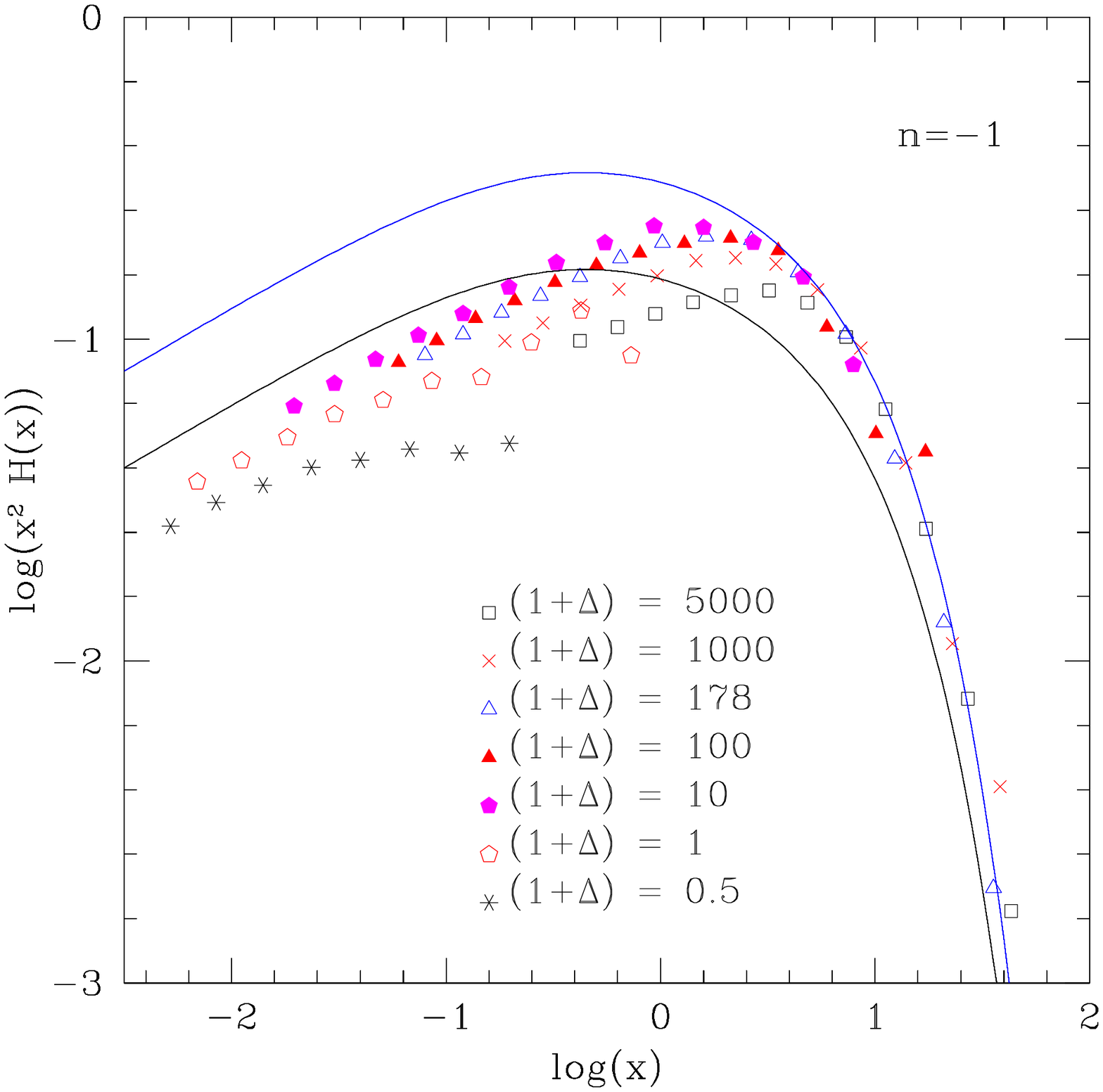,width=8cm,height=6cm}
\centering \psfig{figure=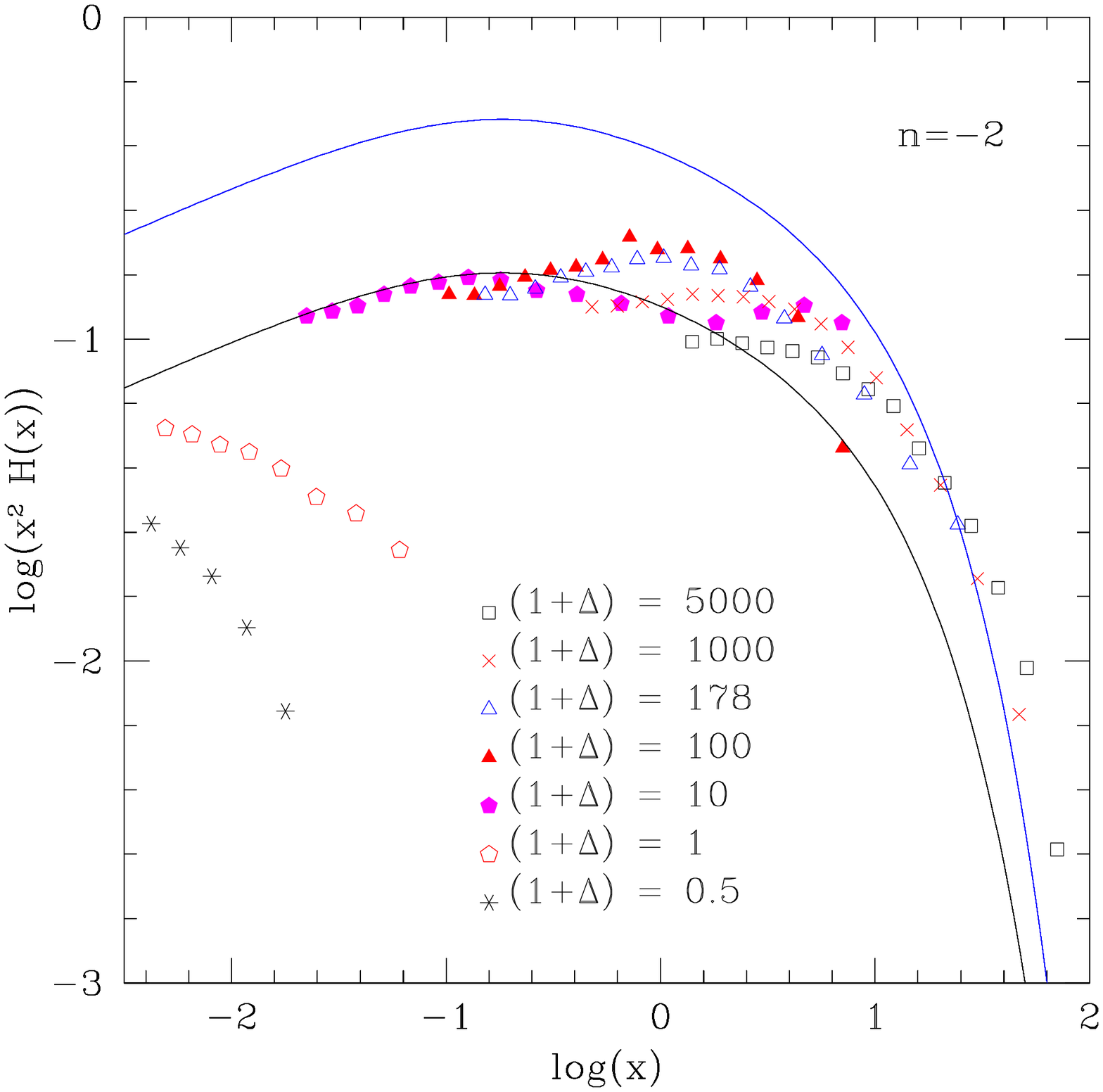,width=8cm,height=6cm}

%{\epsfxsize=8 cm \epsfysize=5.5 cm \epsfbox{figDelfofn0.ps} }
%{\epsfxsize=8 cm \epsfysize=5.5 cm \epsfbox{figDelfofnm1.ps} }
%{\epsfxsize=8 cm \epsfysize=5.5 cm \epsfbox{figDelfofnm2.ps} }

\caption{The mass functions of halos defined by various density
thresholds
$\Delta$ obtained from a ``friend-of-friend'' algorithm. Different
symbols correspond to different values of $(1+\Delta)$. The
data shown correspond to averages over several simulations. The lower
solid curve is the scaling function $h(x)$ measured from counts in
cells while the upper solid curve is $3/\gam \; h(x)$. The non-linear
model predicts that the counts are in-between these two curves at large
$x$, and
below the upper one but generically with the same slope for small $x$.}
\label{figDelfof}

\end{figure}

We show in Fig.\ref{figDelfof} the mass functions we obtain from the
``friend-of-friend'' algorithm. We can check that they are consistent
with the results from the ``spherical overdensity'' algorithm. For
large $\Delta$ we find as previously that the curves superpose beyond
the exponential cutoff but that there is a drift in the power-law
domain.

The advantage of the ``friends-of-friends'' algorithm is that we can
also consider negative density contrasts, providing for much larger a
range for the possible variations of $\Delta$. In such a case we count
one large halo which extends over the whole box (the
clusters linked by the filaments create a percolating network) and
also 
{\it small underdense objects which appear as density peaks within
voids }
 and are separated from the main cluster by their very
low density surroundings. Thus, the mass function defined in this way only
makes sense for very small masses (small parameter $x$). Despite the low density, this is still in the strongly
non-linear regime since  $x \ll (1+\Delta)$ so that $\xia \gg 1$.

\begin{figure}

\centering \psfig{figure=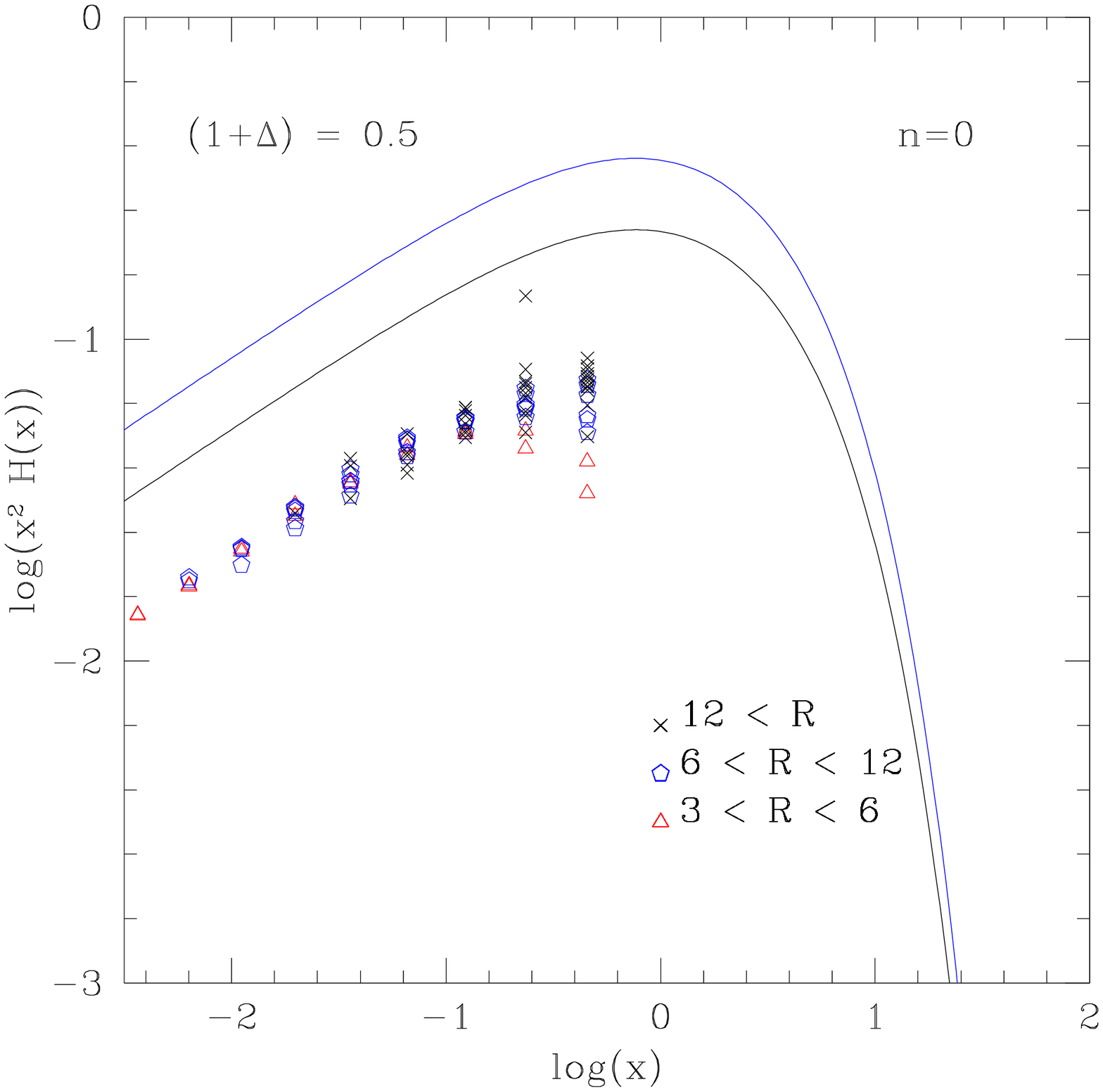,width=8cm,height=6cm}
\centering \psfig{figure=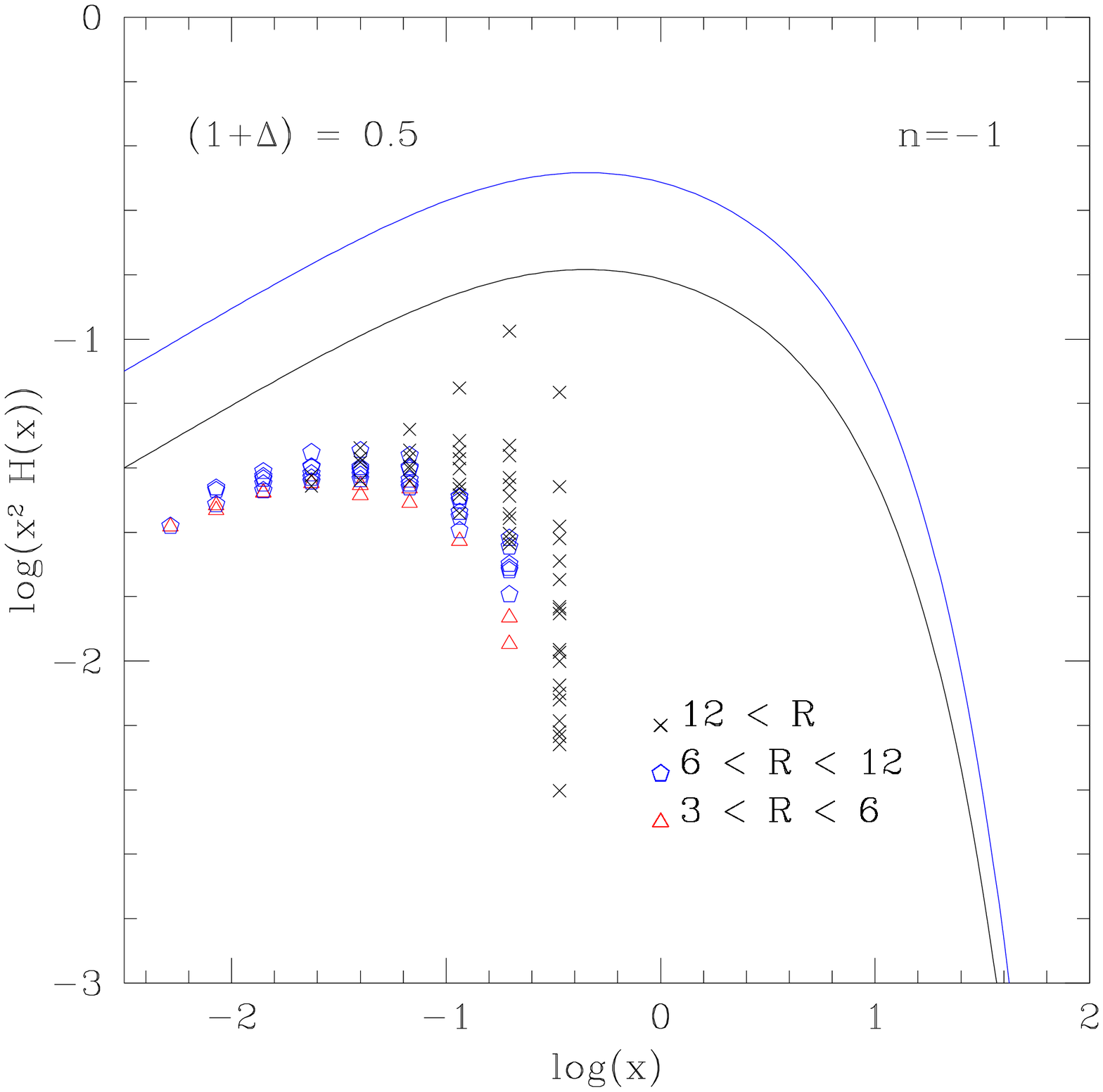,width=8cm,height=6cm}
\centering \psfig{figure=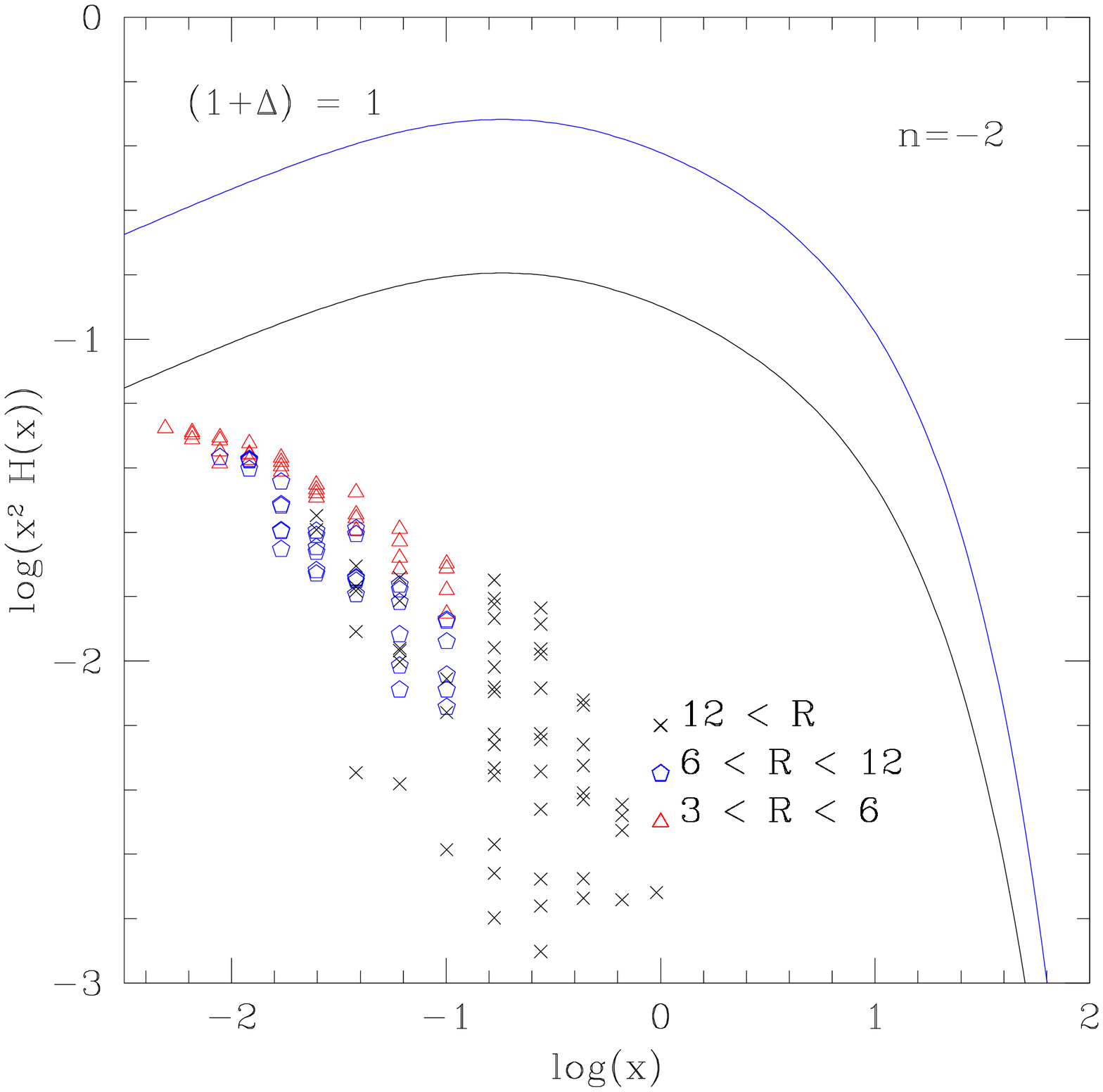,width=8cm,height=6cm}

%{\epsfxsize=8 cm \epsfysize=5.5 cm \epsfbox{figDeltestn0.ps} }
%{\epsfxsize=8 cm \epsfysize=5.5 cm \epsfbox{figDeltestnm1.ps} }
%{\epsfxsize=8 cm \epsfysize=5.5 cm \epsfbox{figDeltestnm2.ps} }

\caption{The mass functions obtained from a ``friend-of-friend''
algorithm for $(1+\Delta)=0.5$ (with $n=0$ and $n=-1$) and
$(1+\Delta)=1$ (with $n=-2$), as in Fig.\ref{figDelfof}. Different
symbols correspond to different ranges for the comoving size $R$ of the
halos. The various points located at the same value of $x$ for a given
range of $R$ correspond to different times (and slightly different $R$
within the allowed range). They should superpose due to the 
self-similarity induced by the power-law initial conditions.}
\label{figDeltest}

\end{figure}

For $n=0$, we can see that all curves fall on top of each other, 
down to
 $(1+\Delta) = 0.5$ which is a little low at $\log x \le -0.5$. 
 Similarly,
for $n=-1$,  the results exhibit the scaling 
 theoretically implied by the stable clustering, except for 
  $(1+\Delta) = 0.5$ at the larger values of $x$ ($\log x \ge -1.5$).
For $n=-2$, also,
the curves with $(1+\Delta) = 0.5$ and $(1+\Delta) = 1$ do not
fall on the scaling curve built by the result at larger density contrasts. As shown in Fig.\ref{figDeltest}, the deviations found for the lower values are seen to be closely associated to the
degradation of the statistics (the counts are low at the larger $x$), 
that occurs for all three values of $n$ we have studied
precisely at the same values of  $(1+\Delta)$ and $\log x$ where the deviations are noticed. Moreover, the deviation at large $x$ of the mass functions measured for low density contrasts ($\Delta \leq 0$) from the scaling predictions corresponds to the fact that they only make sense for small $x$ since we must have $x \ll (1+\Delta)$ so as to avoid entering the linear regime where there are no dense spots within underdense regions like the ones we look for here.

So, the non-linear scaling prediction is seen to hold down to values of
 $(1+\Delta)$ that are unity or even smaller. This is nearly three orders of magnitude below the standard density 
contrast of $178$ usually studied. It has also been tested down 
to $\log(x) = -2.5$, a real improvement  as compared to
existing work: values below $\log(x) = -1$  had never been considered
previously.

Note that the slopes at small $x$ of the mass function 
are close to the ones predicted by the theory. In the $n=0$ case, for instance,
$ \omega \sim 0.5$ , quite close to the slope $\omega = 0.45$ of $h(x)$. 
The offset seen in Fig.\ref{figDelso} with the ``spherical overdensity'' algorithm does not exist in this case.
It should be noted that the $n=0$ case with $(1+\Delta) = 0.5$ clearly shows 
(Fig.\ref{figDelfof})
that the non-linear power-law prediction 
(and perhaps more important, the prediction of existence !) for
 objects that are
much denser than their close environment
is well verified even 
in this quite extreme case of very low density, definitely below the average.

Towards the higher densities, we have run
calculations up to $(1+\Delta) = 5000$,
 a factor of $30$ above the usual densities. As for the spherical density algorithm, scaling is well verified.
Similarly, at the smaller values of $x$, 
steadily increasing deviations appear when $\Delta$
is increased,
although to a lesser extent. For $n=-1$ and $\log(x) = 0$,
the curve for $(1+\Delta) = 5000$  is below the one for
$(1+\Delta) = 1000$  
(that at fixed $x$ corresponds to correlation functions $5$ times smaller, hence to scales $5^{1/1.5} = 2.9$ times larger at fixed time)
by a factor 1.35 and below the $(1+\Delta) = 178$
curve by a factor 1.7. The relevant scales are the smallest reachable in the simulation, typically $R=0.5$ Mpc at $\log(x) =0$ and $R < 0.3$ Mpc at $\log(x) = -0.4$, for $(1+\Delta) = 5000$.
This is to be compared with the force softening parameter which is $0.2$ Mpc in all the calculations presented here. We have checked the sensitivity of the calculations to variations of this softening parameter. Decreasing the latter by a factor of $2$ so as to increase the effect of collisions, lowers the counts by a factor of $1.25$ at the scales considered above
(a change quite comparable to the deviations from scaling that were found),
 but has no effect at the larger values of $x$ (the larger scales). 
For the  calculation  of the correlation
function, at $\xia > 1000$, similarly, deviations of the computation appeared at small scales. 
So, there is a limit towards the large densities and the smaller values of $x$
beyond which
the numerical tests are no longer conclusive, since
  binary collisions due to the discreteness of the points in the simulation
start to
play a non-negligible role.

Although the previous discussion leads to attribute the lack of objects
below some value of $x$, for a given (large) density contrast, to the fact that
the smallest available scale in the simulation has been reached, this
is an important point to be discussed. Note that the theoretical
prediction (see VS) for $H(x)$ extends to all values of $x$ 
(including the lower ones for which there are no numerical data)
and that,
whatever the density contrast, the area under the scaling curve is
unity (it is actually slightly above for $H_{><}(x)$, but is  unity if corrections to avoid double counting are made). This means that however
large the required density threshold is, it defines a way to distribute
all the mass among objects having this density contrast. The
theoretical prediction thus calls for very strong {\it subclustering}
at all density levels. On the other hand, the simulations do not
provide enough power at small $x$ for large overdensities: there is barely
more than $50\%$ of the mass in objects of contrast 5000,
for instance, showing an increasing lack of high density objects.
Again, this apparent difference is worth to be thoroughly investigated
in future more extended simulations, to see whether subclustering up to
extremely high overdensities is present down to very small scales or not.
However, as shown in Valageas (1999), note that the reasonable agreement of the numerical results with our scaling model already shows that there is a large amount of subclustering. It has been steadily argued (Balian \& Schaeffer 1989a,b)
 that the existence and the behaviour of these substructures 
are the base of the scaling model and
govern the properties of the density field.

We have tested the scaling model for the multiplicity function
for vastly different density contrasts, spanning four orders of magnitude,
from $(1+\Delta)=0.5$ to $(1+\Delta)=5000$, finding non-linear objects at every density.
We have shown that up to the small-scale limit where collision effects
begin to play a role and to the low-density limit where deviations from
self-similarity start to appear
the scaling model provides a reasonable approximation to the mass
functions obtained in the simulations.
Also, we can note that the scaling model
gives surprisingly good results in domains which are beyond its
expected range of validity. For instance, the curves for
$(1+\Delta)=10$ and $(1+\Delta)=178$ agree in the numerical results
down to $\xia \sim 1$. Hence, although the model may not be perfect it
is still
quite powerful as it allows one to get a good estimate of the mass
functions of very different objects. Moreover, it clarifies an
interesting link with another statistical analysis of the density
field, namely the counts in cells.

\subsection{Constant radius}
\label{Constant radius}

In practice, one may also be interested in objects which are no longer
defined by a constant density threshold but by a condition of the form
$(1+\Delta) \propto R^{-\beta}$. The predictions of the scaling model
for such a case were studied in VS. 
It
happens that for astrophysical objects like galaxies or Lyman-$\alpha$
clouds the constraints which define these mass condensations can be
approximated in some range by the requirement that their radius is
constant and equal to a given scale which may be associated in the
first case to a cooling radius and in the second to a Jeans length
(see Valageas \& Schaeffer 1999 and Valageas et al.1999 for more
details). This corresponds to the limit $\beta \rightarrow
\infty$. Then, one has to add the supplementary constraint that the
density profile is locally decreasing (since one usually wants to
consider peaks and not valleys). In this specific case one obtains the
relations (see VS):
\beq
H_{cell}(x)=h(x)
\label {HRh}
\eeq
for the simple model. More realistically, one can define (see VS) a scaling function $H_{><}(x)$ for the mass distribution of objects 
of fixed radius and make the statistics of the overdensity $(1+\Delta)$ -hence of the mass since we have $ M \propto (1+\Delta)  $- by requiring for such a halo to have a density contrast larger than $\Delta$ at scale $R$ and smaller than $\Delta$ at a slightly larger scale. Thus one considers individually each object, defined by the fixed radius $R$ and its density contrast $\Delta$ (which gives its mass), making sure that it is surrounded by regions of lower density. Then, one obtains the bounds
\beq
\left\{ \begin{array}{rl} \forall x \; : & H_{><}(x) \leq h(x) \\ \\
x \gg x_s \; : & H_{><}(x) \geq \gam/3 \; h(x) \end{array} \right.
\label {HRbounds}
\eeq
Obviously this procedure is quite similar to the counts in cells so we
can expect the scaling function $H(x)$ to be very close to $h(x)$.

\begin{figure}

\centering \psfig{figure=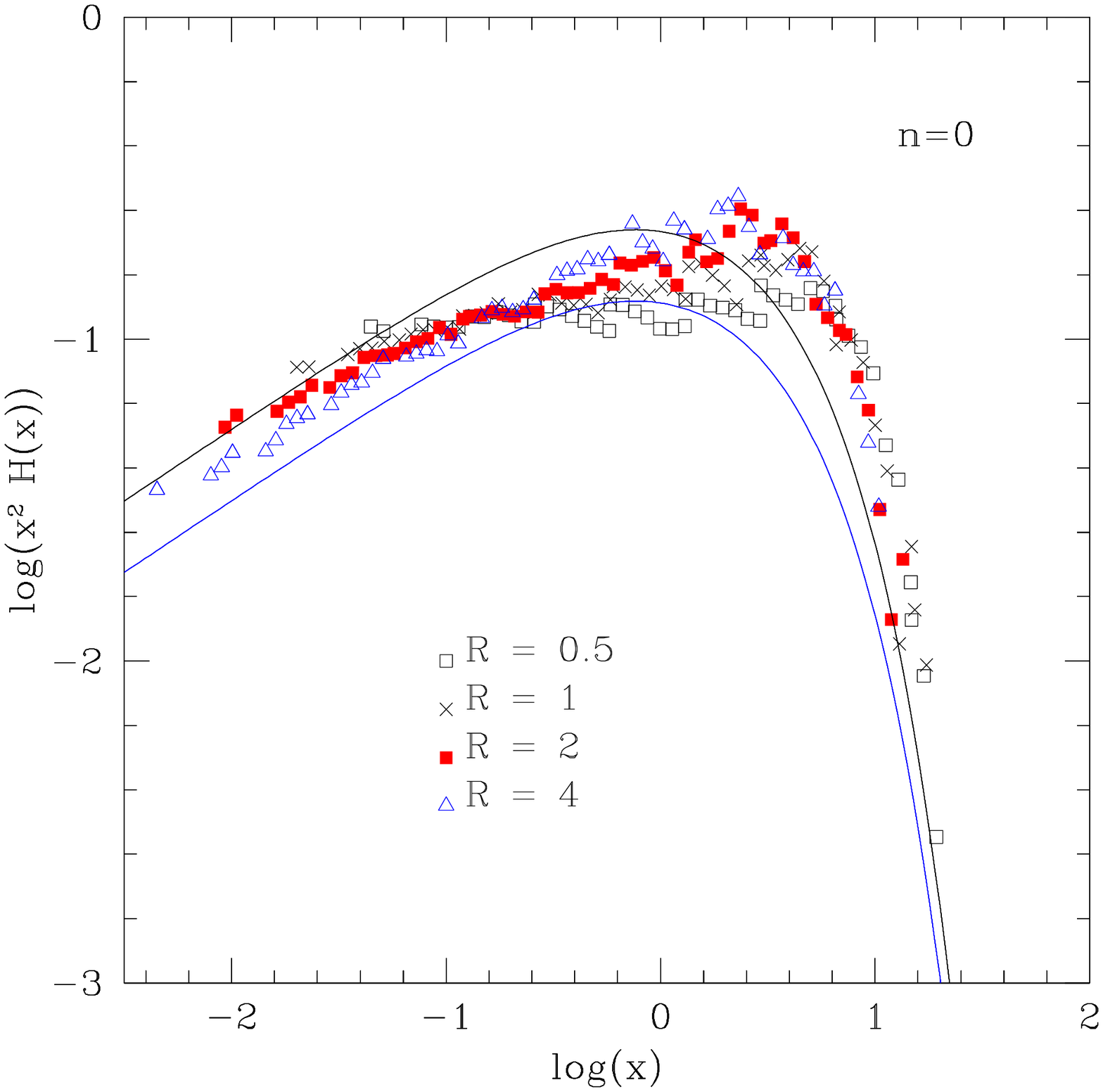,width=8cm,height=6cm}
\centering \psfig{figure=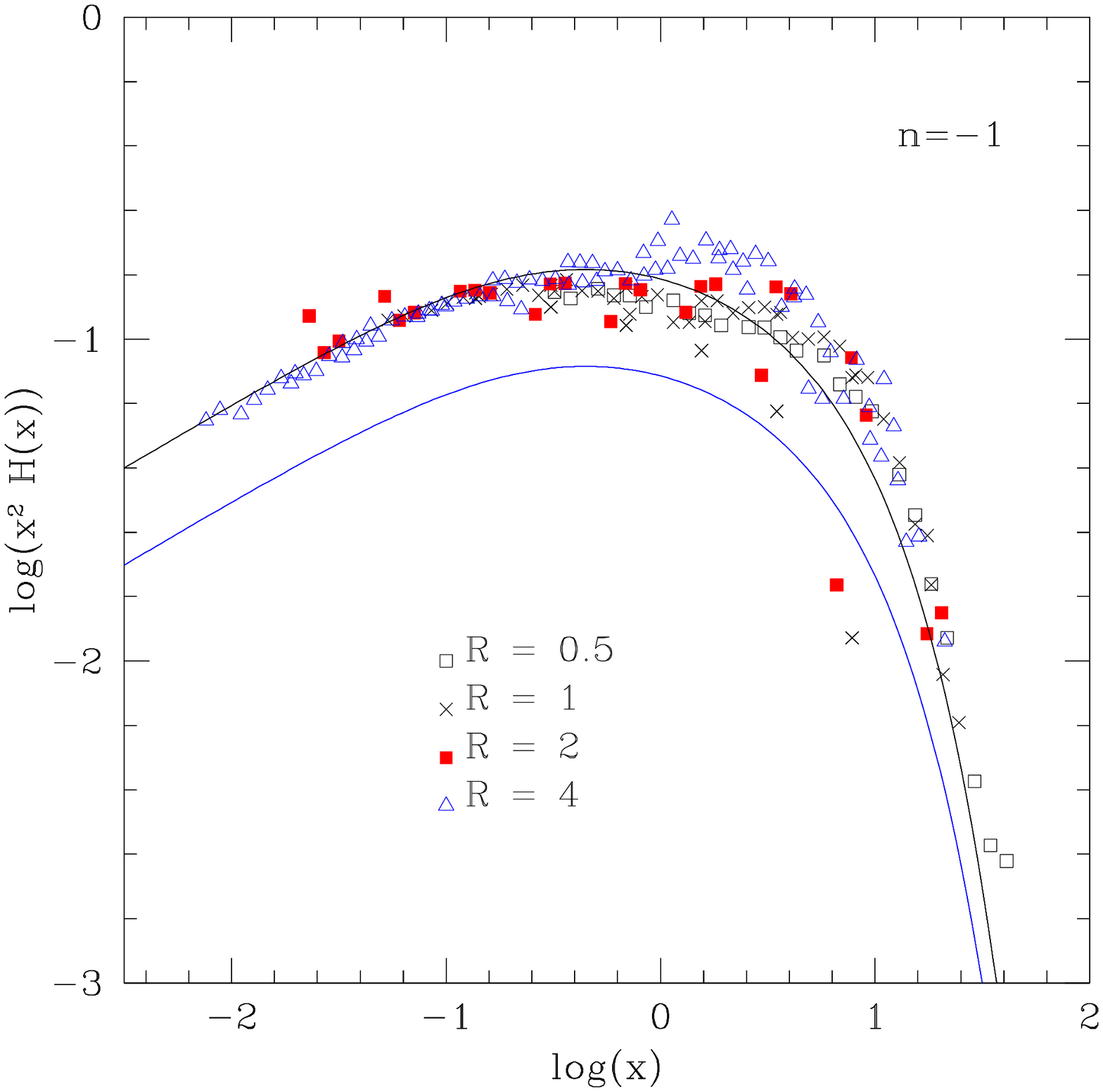,width=8cm,height=6cm}
\centering \psfig{figure=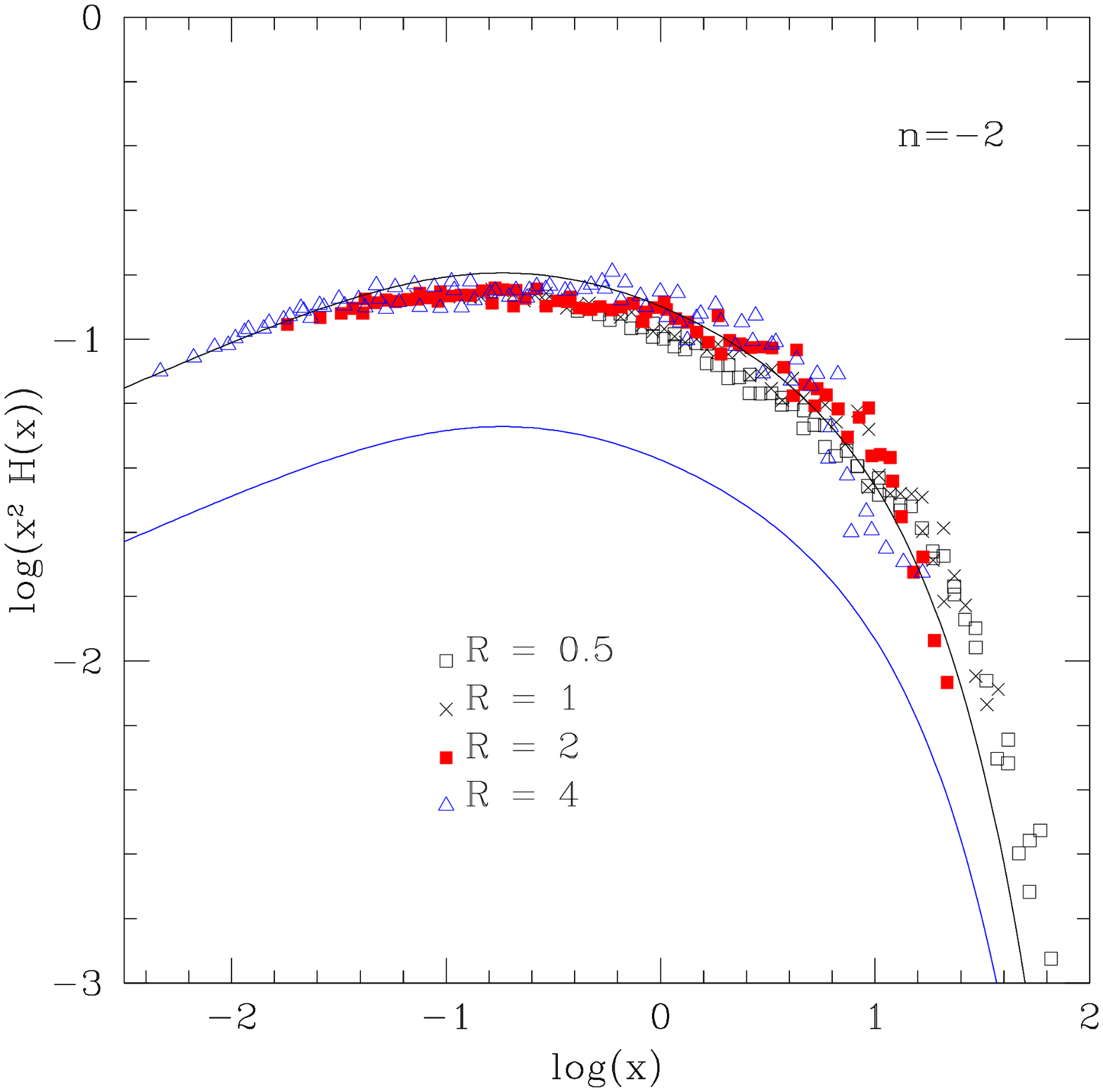,width=8cm,height=6cm}

%{\epsfxsize=8 cm \epsfysize=5.5 cm \epsfbox{figRsoRn0.ps} }
%{\epsfxsize=8 cm \epsfysize=5.5 cm \epsfbox{figRsoRnm1.ps} }
%{\epsfxsize=8 cm \epsfysize=5.5 cm \epsfbox{figRsoRnm2.ps} }

\caption{The mass functions of halos defined by various comoving radii
$R$,
obtained from a ``modified spherical overdensity'' algorithm (see main
text). Different symbols correspond to different values of $R$. Note
that for each radius we display the results obtained at several times
which correspond to various values of $\xia$. The upper solid curve is
the scaling function $h(x)$
measured from counts in cells while the lower solid curve is $\gam/3
\; h(x)$. The non-linear model predicts that the counts are in-between
these two curves at large $x$, and below the upper one but generically
 with the same slope for small $x$.}
\label{figRsoR}

\end{figure}

We show in Fig.\ref{figRsoR} the mass functions obtained in this way from
the numerical simulations. We simply use a modified version of the
``spherical overdensity algorithm'', looking at particles in order of
decreasing density and defining halos as objects of constant size $R$
around density peaks.  The scaling with the variable $x$ is
well verified, for all values of the spectral index $n$, with a result
that is very close to the prediction (\ref {HRh}). There seems,
nevertheless, to be more counts at large $x$ (large masses): the first
of the bounds (\ref{HRbounds}) of the ``improved'' estimate $H_{><}(x)$
(in the sense it is closer to the spherical overdensity procedure)
seems to be somewhat violated. Note however that the curves obtained for different radii (and times) superpose fairly well which shows the predicted scaling in $x$ is verified (within the numerical scatter). 
In fact, this small
deviation is probably due to the definition of $H_{><}(x)$ itself and
not to a failure of the scaling-laws (\ref{scal1}). Indeed, this
estimate of the mass function is derived from specific counts-in-cells
(with the characteristic that these cells consist of an internal sphere
surrounded by a small corona) so that one can miss the most massive
high density halos by looking at a sphere which is not exactly centered
onto the underlying halo. This decreases the mass enclosed in this
sphere, which is maximum for a cell correctly centered. On the other
hand, in the simulation one directly draws the spheres from the highest
density peaks, so that one always counts the largest amount of matter
which can be attached to a given halo (if the object is approximately
spherically symmetric). This latter procedure (looking at the peaks) is not included in
$H_{><}(x)$ and, because of this small ``mis-placement'' of the
spheres, it slightly underestimates the number of very high density and
massive halos. We note that a similar, but very small, deviation
may be seen for the mass functions shown in Fig.\ref{figDelso} and
Fig.\ref{figDelfof}, for the same reason. We also note that the mass function shown in Fig.\ref{figRsoR} allows one to probe deeper into the exponential cutoff of $H(x)$, hence of $h(x)$, as compared to the counts in cells shown in
Fig.\ref{figPN}. This could also be expected from the fact that in this
procedure ``cells'' are drawn directly around highest density peaks,
which are thus well accounted for, while for the statistics of $P(N)$
density peaks can be divided between several cells.

\section{Conclusion}

In this article we have presented a comparison of the results obtained
from numerical simulations to the analytical predictions implied by
the Press-Schechter (Press \& Schechter 1974) prescription and the
scaling model (Balian \& Schaeffer 1989a ; VS).

We have first checked that the two-point correlation functions follow
the behaviour predicted by the stable-clustering ansatz, as shown by
other studies. However, although there is a qualitative agreement
between various works the exact value of the amplitude of the
two-point correlation function in the highly non-linear regime still
remains poorly determined as there remains some discrepancy between
different studies. Next we have shown that the scaling model provides
a good description of the counts in cells statistics. The
characteristic scaling functions $h(x)$ we obtain 
over the unprecedented range $-2.5 < \log x < 1$
agree reasonably
well with previous estimates. Nevertheless, the asymptotic behaviour
of the exponential cutoff is not very well constrained for $n=-2$ due
to the limited range of scales available in the simulations.

Then we considered the mass function of ``just-collapsed'' objects,
which is the quantity most authors have focussed on. As was already noticed
by many studies, we find that the PS approximation works reasonably
well for density contrasts of $\simeq 200$
(although there are some discrepancies). On the other hand, the
numerical results are also consistent with the predictions of the scaling
model (although they do not provide the exact value of the mass
function). Next we have studied more general mass functions defined by
various density contrasts that vary by four orders of magnitude
 (including negative density
thresholds!). The scaling model (which is the only currently available
analytical tool to handle these quantities) was shown to provide a
reasonable estimate of these mass functions.  

Finally, we considered the limiting case of objects defined by a
constant radius constraint (which arises in an astrophysical context
from cooling conditions). We have shown that the scaling model also works
very well for such mass functions.
Some small deviations
appear in certain regimes.
We make specific suggestions on which biases may appear
in the simulations and show they are of the right order of magnitude.
Although these biases likely explain the observed scatter, it is quite
clear that we cannot check the scaling prediction better than this scatter. 
It remains to be verified in more powerful simulations whether 
(at least part of) these deviations can be attributed to some violation
of the scaling predicted by theory, or if these deviations indeed
 disappear and thus
simply reflect the fact that we have pushed the 
present simulation as far as possible, and reached its limits.

Thus, the scaling model provides a reasonable description of
the density field in the non-linear regime (although there are some
deviations in certain regimes they remain reasonably small for
practical purposes). Moreover, it allows one to link two different
properties of the latter: the counts in cells statistics and the mass
functions. In addition, it can be used to obtain many different mass
functions (in addition to the usual ``just-collapsed''halos) which is
of great interest for practical purposes when one intends to model
various objects like Lyman-$\alpha$ clouds or galaxies which clearly
cannot be defined by the sole constraint $\Delta = 177$. Obviously,
our study should be extended to more realistic power-spectra which are
no longer power-laws (e.g. CDM) and to different cosmologies
(low-density universes).

\end{document}